\documentclass[a4paper,fleqn,usenatbib]{mnras}

\usepackage{amsmath}	

\usepackage{txfonts}

\usepackage[T1]{fontenc}
\usepackage{ae,aecompl}

\usepackage{graphicx}	
\usepackage{amsmath}	
\usepackage{amssymb}	
\usepackage{hyperref}

\usepackage{color} 
\usepackage{subcaption}
\usepackage[algosection, boxruled, linesnumbered]{algorithm2e}
\SetProcNameSty{textsc}
\SetProcArgSty{textsc}

\makeatletter
\newcommand{\removelatexerror}{\let\@latex@error\@gobble}
\makeatother

\title[Linear feature detection algorithm]{Linear feature detection algorithm for astronomical surveys - I. Algorithm description}

\author[D. Bekte\v{s}evi\'{c} and D. Vinkovi\'{c}]{
Dino Bekte\v{s}evi\'{c},$^{1,2}$\thanks{E-mail: dino@iszd.hr (DB)}
and Dejan Vinkovi\'{c}$^{1,2,3}$
\\
$^{1}$Faculty of Science, University of Split, Rudjera Bo\v{s}kovi\'{c}a 33, HR-21000 Split, Croatia\\
$^{2}$Science and Society Synergy Institute, Bana Josipa Jela\v{c}i\'{c}a 22, HR-40000 \v{C}akovec, Croatia\\
$^{3}$HiperSfera d.o.o., Ilica 36, HR-10000 Zagreb, Croatia
}

\date{Accepted XXX. Received YYY; in original form ZZZ}

\pubyear{2017}

\begin{document}
\label{firstpage}
\pagerange{\pageref{firstpage}--\pageref{lastpage}}
\maketitle

\begin{abstract}
Computer vision algorithms are powerful tools in astronomical image analyses, especially when automation of object detection and extraction is required. Modern object detection algorithms in astronomy are oriented towards detection of stars and galaxies, ignoring completely detection of existing linear features. With the emergence of wide-field sky surveys, linear features attract scientific interest as possible trails of fast flybys of near-Earth asteroids and meteors. In this work we describe a new linear feature detection algorithm designed specifically for implementation in Big Data astronomy. The algorithm combines a series of algorithmic steps that first remove other objects (stars, galaxies) from the image and then enhance the line to enable more efficient line detection with the Hough algorithm. The rate of false positives is greatly reduced thanks to a step that replaces possible line segments with rectangles and then compares lines fitted to the rectangles with the lines obtained directly from the image. The speed of the algorithm and its applicability in astronomical surveys are also discussed.
\end{abstract}

\begin{keywords}
methods: data analysis -- surveys -- meteors -- minor planets, asteroids: general
\end{keywords}



\section{Introduction}

Long linear features in astronomical images are seldom of scientific interest, with the exception of asteroid tracks \citep[in case of fast near-Earth-crossers, but they are still much shorter than the meteor and satellite tracks;][]{asteroids}. Such features are treated as noise, since the expectation is that they originate from non-astronomical sources: satellite and airplane tracks, scratches and dirt within the optical system and imaging camera, reflection and diffraction spikes within the telescope \citep{Storkey}. With the growing importance of sky surveys of various size and scope \citep{Surveys}, it also grows the need for a proper classification of all objects detected within the field of view. This also includes linear features, as their identification enables corrections to the scientifically relevant features affected by the linear noise.

However, the large volume of imaging data produced in surveys, especially in the wide field surveys in visual bands, created two interesting challenges. The first is that detection of linear features has to be done automatically, with a robust linear feature detection algorithm (LFDA). The second is that some specific cases of linear features attracted scientific interest: trails crated by close flybys of the near-Earth asteroids (NEAs) \citep{SDSSMovingObjects,LSSTCam} and, more recently, trails created by meteors \citep{BiDS14,Vinkovic}. Those are rare transient events, with unpredictable positions in the sky (except when a NEA flyby is predicted far in advance, but most of the time smaller NEAs are discovered during their close flyby). Wide field sky surveys cover a large fraction of the sky, with modern surveys doing that repeatedly. This creates a big enough sky coverage to detect such rare random events, but they are hidden in a large volume of imaging data.

Detection of lines is a very common requirement in computer vision. The standard method used for this is the Hough algorithm \citep{Hough} or its derivatives. Hence, the first suggestions for LFDA in astronomy were based on this algorithm \citep[e.g.][]{Kubickova,Storkey}, but detecting lines in astronomical images is not the same class of problems as detecting lines in ordinary photos. Astronomical images have stars and galaxies that produce strong localized brightness peaks that easily confuse the Hough algorithm. This manifests itself as a large number of false positives, which forces the algorithm to be limited only to very bright lines that dominate the image. This ignores all the false negatives coming from lines of lower brightness.

In this paper we describe a new type of LFDA that overcomes the above mentioned problems. The Hough method is still one of the crucial elements, but we added several other steps that expand the applicability of the line detection. For the survey images we used the SDSS database \citep{SDSSDR12}. A detailed review of our LFDA performance on the entire SDSS image dataset will be published in an upcoming paper, while here we focus solely on the algorithm description and its basic performance properties. In Section \ref{sec:limitations} we first describe the constraints on the applicability of any LFDA in Big Data astronomy. An overview of the algorithm and its algorithmic components are described in Section \ref{sec:overview}. The first step in our LFDA is removal of all objects (stars, galaxies, etc.) from the image, such that the linear features can dominate in the Hough space. This step depends on the particular survey under consideration, while we describe in Section \ref{sec:objectremoval} how we do it in the case of SDSS. Steps in our LFDA differ slightly between detection of bright lines, described in Section \ref{sec:DetectBright}, and detection of dim lines, described in Section \ref{sec:DetectDim}. In Section \ref{sec:Examples} we show some typical examples of line detection, including intermediate steps of the algorithm performance. In Section \ref{sec:DiscussionConclusion} we discuss the algorithm's performance in the context of false positive and false negative detections and the algorithm's implementation.

\section{Algorithm applicability and limitations.}
\label{sec:limitations}

The LFDA is not particularly useful in situations when images can be inspected manually for linear features. If we assume that the manual revision time for one image is one second then a set of 10 000 images would require approximately 3 hours for such a task, while 100 000 images would take about a day of revision time. However, the total revision time is mainly dominated by the fatigue of the reviewer prolonging the total review time, making it a very laborious process. Thus, we can stipulate that there would be no need to implement a computer vision algorithm to deal with a line detection problem in up to 100 000 images. The number is even lower if we need to inspect images multiple times.

This limits the algorithm use cases to situations where we need to analyze images in real time or when we need to inspect image sets with the number of images in the millions. The former use case is well represented by potential applications in the large sky surveys that require real time image analysis, such as the Large Synoptic Survey Telescope (LSST)\footnote{\url{http://www.lsst.org/}}, while the latter is applicable in cases of existing image databases, such as the Sloan Digital Sky Survey (SDSS)\footnote{\url{http://www.sdss.org/}} or the Dark Energy Survey (DES)\footnote{\url{http://www.darkenergysurvey.org/}}.

In both use cases it is necessary for the algorithm to perform fast. For example, the SDSS database contains approximately 6.5 million images\footnote{\url{http://www.sdss.org/dr13/data_access/volume/}}. If it takes a second for the algorithm to decide if an image contains a linear feature or not, it would take a total of 75 days to process the entire data set. It is therefore desirable that the total temporal execution per image is less than a second, including data acquisition, not just the processing time. Also, the algorithm should be easily parallelizable to speed up the analysis of the entire database. Analysis of an image is not dependent on any of its neighbouring images. As long as there is no shared input or output objects or resources that can not be shared simultaneously, the program parallelization becomes an embarrassingly parallel problem. Hence, maintaining low data access times is a greater problem than parallelization, except in a case of real-time execution when the images come directly from the detector.

If the image storage facilities are not equipped to run the detection algorithm on-site, it is necessary to consider the computer memory required to store the data on a machine that would be capable of running the algorithm. For example, the corrected SDSS bzipped FITS files require 15.37TB of storage. Once bunzipped for processing the total memory requirement rises to 70-80TB, assuming the average file size of 13MB. This example is the lower estimate of the total memory requirements as the SDSS camera has ``only'' 126 megapixels \citep{Gunn98}, compared to more modern cameras such as the 560 megapixels DECam \citep{DECam}, or 3 200 megapixel LSST camera \citep{LSSTCam}.  Additionally, if the processing could not be done on-site the question of how to transfer such a large quantity of data to the machine becomes important. With the hard disk drive (HDD) storage requirements only growing, it is obvious that the processing will eventually have to be done in real time, within the image processing pipeline of the sky survey in question, or on compressed sections of the total data that can fit into the machine where the processing will be done. In the latter case, uncompressing the data puts additional strains on the temporal execution requirements of the algorithm itself. Our test on bz2 compressed SDSS FITS files, with 100 test samples performed on an i5 Intel Asus laptop with 7 200 rpm Seagate Momentus HDD with an average read rate of 78.6 MB/s and 15.76 ms access time, shows consistent and stable bunzip execution times of 0.8 seconds.

Less worrying than the HDD memory requirements there are also random access memory (RAM) requirements. During algorithm execution each of the images might require several copies to be held in memory during processing. If several LFDA processes depend on the same RAM space, and each holds several copies of an image, then the number of processes that are able to run in parallel could be very limited. Fortunately, RAM constraints are subverted by the fact that images themselves are not that large. In the case of SDSS the image frames are 2048x1489 pixel arrays, while the DES images have 2000x4000 pixels. This limits image size to the range of 10-30MB, which means that even in a case when a single process requires 100-120MB it would be possible to run up to ten processes per 1GB of RAM space.

The last constraint on the algorithm execution is its correctness. For example, when we tried to analyze 6.5 million image frames from the SDSS database, it was crucial to minimize Type I (false positive) and Type II (false negative) error rates. If we assume that the software rejects 99\% of images, as images without linear features, the number of returned images, possibly containing linear features, would be approximately 65 000 images. Each one of those would have to be reviewed manually to determine if it is a false positive or not. On the other hand, severely restricting detection parameters to accept only the most confident of detections would boost Type II errors and result in a biased sample of only the most bright, longest and thickest linear features and would therefore be a poor statistical sample unfit for making general conclusions.

\section{Algorithm overview}
\label{sec:overview}

\subsection{Platform and data specifics of our LFDA implementation}
\label{subsec:PlatfromSpecific}

We implemented our LFDA algorithm in Python 2.6.6\footnote{\url{https://www.python.org/}} and utilized Erin Sheldon's fitsio v0.9.4\footnote{\url{https://github.com/esheldon/fitsio}}, \emph{numpy} v1.7.0\footnote{\url{http://www.numpy.org/}} and \emph{OpenCV} v2.4.8\footnote{\url{http://opencv.org}} for reading/writing FITS files and image analysis. In order to maintain the fast performance requirements, some steps (such as object removal, data input and congregation of results) are coded to give optimal performance for the dataset in question. Currently the algorithm is specifically set up for the SDSS data structure and for execution on the Fermi cluster at the Observatory Belgrade. However, a lot of attention has been given to abstracting the data layer input-output operations, which makes the algorithm easily adaptable to other types of datasets and platforms. The existing software stack contains seven modules in total. Most of them deal with input, output, various algorithm settings and will not be discussed in depth in this paper. We will focus only on the modules that contain the gist of the linear feature detection algorithm.

Distributed Queueing System\footnote{\url{http://www.csb.yale.edu/userguides/sysresource/batch/doc/InstMaint.html\#_Toc352941915}} (DQS) scripts control the process requirements (number of CPU cores, memory, file requests, etc.) and parallelized execution by instructing the TORQUE\footnote{\url{http://www.adaptivecomputing.com/products/open-source/torque/}} resource manager and Maui\footnote{\url{http://www.adaptivecomputing.com/products/open-source/maui/}} scheduler on the requirements of each process within a submitted parallel job. These scripts are generated by the {\it Createjobs} module (see pseudo-code \ref{alg:createjobs}) and, apart from the intrinsic process requirements, consist mostly of calls made to the {\it Detecttrails} module (see pseudo-code \ref{alg:detecttrails}). Each job is given a subset of images for analysis, disjunct from the other jobs. Within an individual job, images are processed in serial order and the job has its own isolated output file. Parallelization, thus, is dependent on the environment in which the code is running and it is not inherent to the actual line-detection algorithm. The {\it Detecttrails} module contains all of the 29 parameters that control the algorithm execution, listed in Table \ref{tb:detpar}, and invokes the commands in the {\it Removestars} and {\it Processfield} modules in the correct order. The {\it Removestars} module contains the known object-removal functions (stars, galaxies, etc.) while {\it Processfield} module is where the detect bright and dim functions are located.

\begin{procedure}
  \SetKwInOut{Input}{Input}
  \SetKwInOut{Output}{Output}
  \Input{Execution and detection parameters, execution limitations, execution environmental variables, }
  \Output{Series of DQS scripts and a batch shell script.}

  read DQS file template\;
  \While{not out of parameters}{
    replace KEYWORDS\;}

    save DQS scripts as job\#.dqs\;
    open new file\;
    \ForEach{DQS script}{
      write ``submit job\#.dqs''\;}

    save as shell script\;
\caption{Createjobs()}
\label{alg:createjobs}
\end{procedure}

\begin{procedure}
  \SetKwInOut{Input}{Input}
  \SetKwInOut{Output}{Output}
  \Input{Detection parameters, set of files to process.}
  \Output{Results (text file) or none.}

  \ForEach{file in set}{
    bunzip file to BUNZIP\_PATH\;
    open image header with fitsio\;
    open data header with fitsio\;
    call remove\_stars in RemoveStars module\;
    call detect\_bright in Processfield module\;
    \uIf{detection}{
      write results\;
      continue\;
      }
    \Else{
      call detect\_dim in Processfield module\;
      \uIf{detection}{
        write results\;
        continue\;
      }
      \uElseIf{error}{
        write errors\;
        continue\;
      }
      \Else{
        continue\;
      }
    }
   }

\caption{DetectTrails(). Called from DQS scripts.}
\label{alg:detecttrails}
\end{procedure}

\subsection{General algorithm flow}
\label{subsec:GeneralFlow}

The algorithm itself is designed as a three-step pipeline in which each step is more time consuming and less reliable. A short diagram of the execution logic is shown in the pseudo-codes \ref{alg:detecttrails} and \ref{alg:createjobs}. A more detailed algorithm-execution flowchart is shown in Figure \ref{fig:DetailedFlowChart} (Appendix \ref{sec:detailedalgflow}). The first step is the removal of all known objects, followed by a step that tries to detect bright linear features. When this step detects a possible linear feature, results are outputted and the process terminates. If there are no bright linear features detected then the next step is to search for dim linear features. The image is subjected to additional processing that increases the intensity of dim objects, followed by a step similar to detecting bright lines. Detecting bright and dim lines includes several checks to test the validity of the possible linear features. Whenever any of the checks fail, the algorithm immediately rejects the line candidate without completing further checks.

The central idea of the algorithm is to remove all known objects in the image, detect the remaining objects, apply shape descriptor operators to perform structural analysis on the detected remaining objects, and then reconstruct the image while keeping only the elongated objects. If the same object is detected in the reconstructed image and in the image with known objects removed, then the detection can be confidently declared as positive. This approach is used in both the detect bright lines and the detect dim lines steps. Such a two-step search based on line brightness allows the algorithm to be more robust against varying image quality.

\subsection{Erosion and dilation}
\label{subsec:ErosionDilatation}

\begin{figure}
\begin{subfigure}{.49\textwidth}
  \centering
  \includegraphics[width=0.9\linewidth]{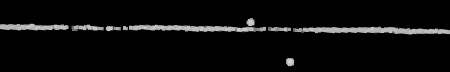}
  \caption{Dilation}
  \label{fig:dilatio1}
\end{subfigure}
\begin{subfigure}{.49\textwidth}
  \centering
  \includegraphics[width=.9\linewidth]{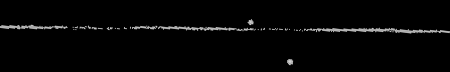}
  \caption{Original}
  \label{fig:original1}
\end{subfigure}
\begin{subfigure}{.49\textwidth}
  \centering
  \includegraphics[width=.9\linewidth]{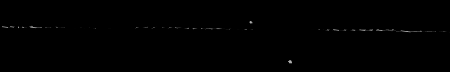}
  \caption{Erosion}
  \label{fig:erosion1}
\end{subfigure}
\caption{An example of the effects dilation and erosion have on linear objects. This original image is a part of the SDSS image frame {\it frame-i-002888-1-0139.fits}. A 3x3 rectangle kernel was applied in both operations with all kernel elements used for the decision on the new value for the anchor point.}
\label{fig:morph_op}
\end{figure}

\begin{figure}
\begin{subfigure}{.49\textwidth}
  \centering
  \includegraphics[width=0.9\linewidth]{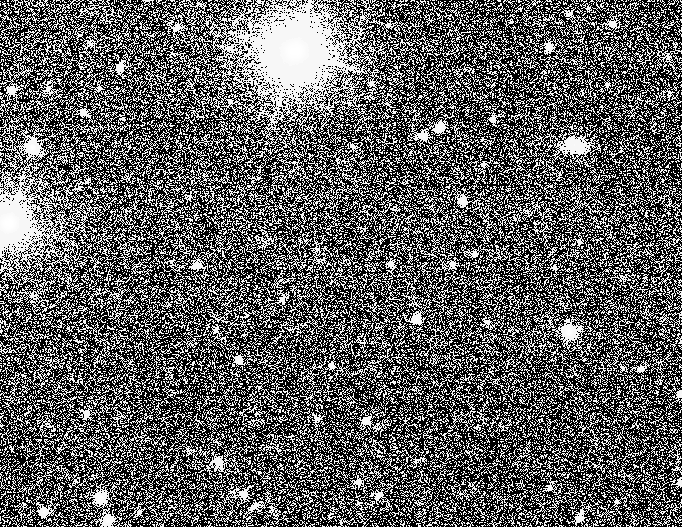}
  \caption{Original image.}
  \label{fig:original2}
\end{subfigure}
\begin{subfigure}{.49\textwidth}
  \centering
  \includegraphics[width=.9\linewidth]{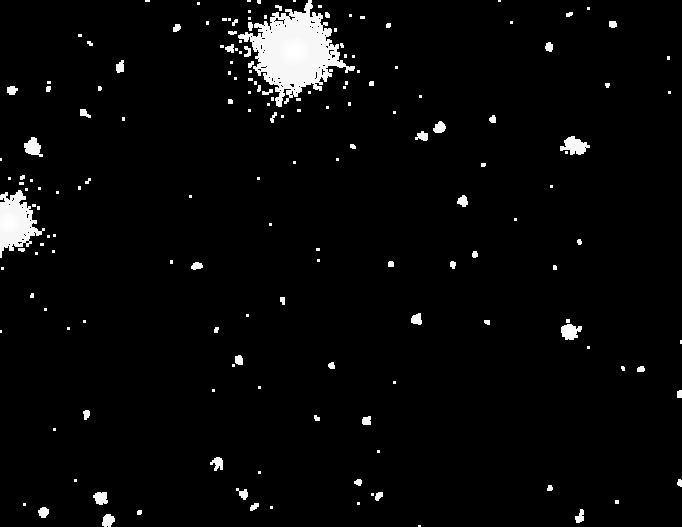}
  \caption{Original image after opening operator has been applied.}
  \label{fig:opening}
\end{subfigure}
\caption{Opening operator (erosion followed by dilation) is very useful at removing noise, as shown on this portion of the SDSS image frame {\it frame-i-002888-1-0017.fits}. A 3x3 rectangle kernel was used.}
\label{fig:noiserem}
\end{figure}

Erosion and dilatation are morphological operators used in the LFDA for noise removal. Both erosion and dilation use a kernel applied to the image pixels. The kernel is a matrix with a defined anchoring element, usually at its center. A new value of the element located at the anchor point is decided by the image pixel values covered by the nonzero elements of the kernel. Erosion will assign the anchored pixel the minimum value found in the kernel, while dilatation will assign the maximum value found in the kernel.

Chaining combinations of erosion and dilation operators creates other operators. The opening operator is erosion followed by dilation. As we show in Figure \ref{fig:morph_op}, applying a too-large erosion operator has the potential of destroying objects in the image. However, if just sufficiently strong erosion is applied, such that the object of interest survives, and then equally strong dilation is applied, the resulting image would contain the original object of interest and all objects larger than it, but no objects smaller than the object of interest. The opening operator is thus very good at dealing with noise. A small erosion kernel effectively deletes isolated pixels, after which the dilation operator restores remaining objects in the image to their previous sizes. This process is shown in Figure \ref{fig:noiserem}. The opening operator tends to break up diffuse objects, such as halos of large stars and galaxies, into a collection of small "blobs". To reconnect these blobs back into a singular object, it is necessary to apply dilation with a larger kernel than the one used in erosion.

\subsection{Histogram equalization}
\label{subsec:HistogramEqualization}

\begin{figure}
\begin{subfigure}{.49\textwidth}
  \centering
  \includegraphics[width=0.9\linewidth]{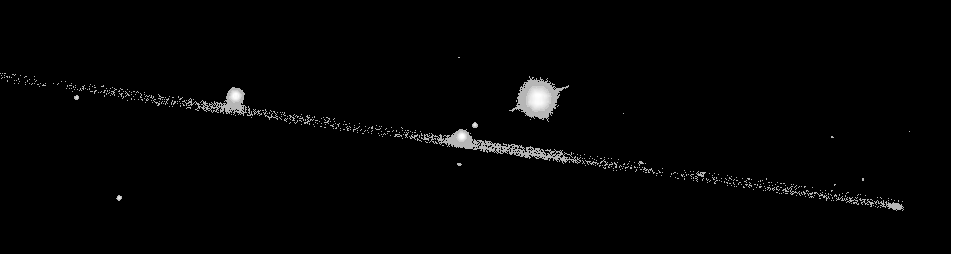}
  \caption{Only histogram equalization.}
  \label{fig:heqDM}
\end{subfigure}
\begin{subfigure}{.49\textwidth}
  \centering
  \includegraphics[width=.9\linewidth]{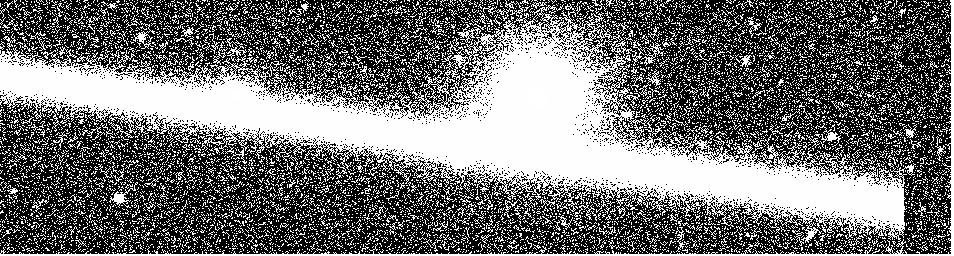}
  \caption{Brightness was increased by 0.5 before histogram equalization.}
  \label{fig:heqBR}
\end{subfigure}
\caption{An example of histogram equalization (adjusting image contrast) without and with brightness enhancement. This technique is used for enhancing the visibility of existing linear dim features. The image is a part of SDSS image frame {\it frame-i-005973-3-0130.fits}.}
\label{fig:heq}
\end{figure}

Histogram equalization is a technique used for adjusting image contrast. In essence, histogram equalization finds the smallest and largest intensity values in the image, assigns them the minimum and maximum allowed intensity value (set by the image bit depth) and then reassigns the intensity values of the remaining pixels based on a cumulative distribution function of the image. In our LFDA, histogram equalization is applied to images in conjunction with brightness enhancement. Brightness enhancement consists of manually adding a fixed value to all existing pixels with intensities larger than zero. Increasing the brightness and then applying histogram equalization can lead to drastically different end results depending on the amount of brightness added (see Figure \ref{fig:heq}). Increasing brightness and contrast of an image increases the amount of noise in the image as well.

\subsection{Canny edge detection}
\label{subsec:CannyEdge}

Canny edge detection is a method of finding edges in a picture. In our LFDA, it is used for detecting minimum area rectangles (see Figure \ref{fig:DetailedFlowChart}). It was first developed by \citet{Canny}, but today there are several variations of the algorithm. The version used in our LFDA is the OpenCv implementation. First the image is blurred with a 5x5 Gaussian blur matrix to reduce the effects of remaining noise. Then the Sobel operator \citep{Sobel} is applied. It is based on two 3x3 kernels, $G_x$ and $G_y$, which are used for calculating the approximate derivatives of the intensity gradient at a pixel in the horizontal and vertical directions respectively. A new image representing the gradient magnitude is constructed from the two images as
\begin{equation}
 G = \sqrt{G_x^2 + G_y^2}.
\end{equation}
As the gradient will be the largest at the pixels with the largest change in intensity, the end result is an image with emphasized edges. The gradient direction can be calculated as:
\begin{equation}
 \Theta = \arctan\frac{G_y}{G_x}
\end{equation}
Edge direction is perpendicular to the gradient direction, and is rounded to be in one of 4  possible directions: 2 diagonal, vertical or horizontal direction.

To reduce the number of falsely detected edges a non-maximum suppression is applied. Every nonzero gradient point (i.e. every nonzero pixel in the constructed image) is checked and compared to the neighboring gradient points, in the respective gradient directions, to see if the gradient point in question is the local maximum or not. If that particular point is not the local maximum, the gradient at this point is set to zero. This is an edge thinning technique, as the resulting image represents an image containing only the points of sharpest change in intensity value, while all other pixels are removed. The remaining edges, are filtered through two user defined thresholds (lower and higher) to remove noise-induced edges. This is known as hysteresis thresholding. Pixels that are above the higher threshold are automatically declared to be actual edges. Pixels bellow the lower threshold are automatically discarded. The remaining pixels between lower and higher thresholds are judged by the connectivity criterion. If such a pixel can be connected through any number of steps to a pixel above the higher threshold it is declared an edge, otherwise it is discarded. The result is a binary image containing edges of the input image.

\subsection{Contour detection}
\label{subsec:ContoursDetection}

A contour is an outline representing the shape of an object. We use a contour-detection method first developed by \citet{Suzuki85}, which takes detected edges and finds contours of objects. Unlike the Canny edge-detection algorithm the result of contour detection is a list of 2D points (vectors). The function we use is taken from OpenCV, and it has two important parameters - \emph{mode} and \emph{method}. The \emph{mode} parameter describes relationships between the contours that belong to the same object. Considering that the main interest is the shape of an object, it is necessary to retain the most outer object contour. However, to avoid problems with edges near the image borders and poorly detected edges, it is prudent to consider returning all detected contours and examining them as well. This is the currently used mode in our LFDA.

The \emph{method} parameter controls contour-detection approximations and deals with the way the contours are stored. Since these contours will be used only for fitting minimum-area rectangles, it would be enough to use a simple approximation where all horizontal, diagonal and vertical segments are compressed and only their end points are stored. However, as with the \emph{mode} parameter, it is again prudent to return all contour points and use a more reliable shape descriptor to select contours of interest for linear feature detection.

\subsection{Minimum area rectangle}
\label{subsec:MinimumAreaRectangle}

Minimum-area rectangles are rectangles of arbitrary rotation, but with the minimum surface area that still encompasses the entire contour. A fast {\it O}(n) method for fitting minimum area rectangles using contours was first proposed by \citet{MinAreaRect}, based on the previous work by \citet{FreemanShapira}. The algorithm exploits the fact that a rectangle of minimum area enclosing a convex polygon has a side collinear with that one of the edges of the polygon. The first step in the algorithm is to find the object's vertices of minimum and maximum x and y coordinates. Then parallel lines are drawn through the vertices, representing the first calipers. The second caliper lines are drawn perpendicular to the first and touching the object at its extreme vertices in the opposite direction. These calipers are rotated around the object's contour and the enclosed area is recalculated and compared until the rectangle with the minimum area is found. The benefiting factor is that the rotation does not need to be incremental because, as given by \citet{FreemanShapira} theorem, one of the sides of the calipers must be collinear with one of the sides of the object and therefore it is sufficient to rotate the calipers by the minimum angle one of the caliper hands closes with the contour.

The main motivation for finding the minimum area rectangles from contours is that we are able to gain descriptive information about the contours (e.g., lengths of the rectangle sides). Additionally, this approach has proven itself to be more reliable than selecting a simple contours method in conjunction with the contours \emph{mode} parameter set to return only the outermost contours. By requiring that the ratio of the rectangle's side lengths differs from 1, it is possible to find only those contours that encompass elongated objects in the image.

\subsection{Hough transform}
\label{subsec:HoughTransform}

The generalized Hough transform refers to a feature-extraction technique used in computer vision that is based on a voting procedure that favors a certain class of shapes. The classical Hough transform \citep{Hough} deals with detecting linear shapes in images. In Cartesian coordinates a line can be represented by an equation
\begin{equation}\label{eq:y1}
    y=mx+b.
\end{equation}
transforming equation \ref{eq:y1} into the normal-line parameterization \citep{Hart09}, we obtain
\begin{equation}\label{eq:y2}
    y= - \frac{\cos(\theta)}{\sin(\theta)}x+\frac{r}{\sin(\theta)}.
\end{equation}
Now we can rewrite equation (\ref{eq:y2}) into the Hessian normal form
\begin{equation}\label{eq:r}
    r = x\cos(\theta) + y\sin(\theta).
\end{equation}
The transformed coordinate space which has $\theta$ and $r$ as its axes is called the Hough space. A set of points belonging to a line in Cartesian space will get mapped to a set of sinusoids intersecting at a point $(\theta, r)$ in the Hough space. Thus, we reduced the problem of detecting a line in an image to a problem of detecting points in Hough space. Once the local maximum in Hough space is detected, an inverse transform can be performed using equation (\ref{eq:y2}) to get the corresponding line parameters in the Cartesian space.

The basic Hough transform is fairly easy to reproduce. First set a threshold value that determines which pixels on an image are considered for line detection (i.e., pixels with intensity greater than the threshold). Then for each active pixel $(x, y)$, draw a family of lines $(\theta_i, r_i)$ in the Hough space. Each consecutive line in the family has its $\theta_i$ increased by a preset value defined by the resolution of Hough space. For each $\theta_i$ calculate $r_i$ and increase the vote in the Hough-space accumulator at $(r_i, \theta_i)$. Once all active pixels have been exhausted, search for the global maximum in Hough space and declare these Hough space coordinates to be the detected line parameters. A search for local maxima would produce a set of lines found in the image. Additionally, only longer lines can be selected by setting a lower limit for votes in the Hough space. However, a lot of subtlety is hidden in the way these local maxima are found. In Appendix \ref{app:HoughEx} we illustrate why accumulator maxima have to be selected carefully, how Hugh transform can produce false positives, and why any disk-like object (stars, galaxies, nebulae, etc.) has to be "hollowed" out or completely removed in order to prevent false line detections.

\begin{figure}
  \centering
  \includegraphics[width=0.49\textwidth]{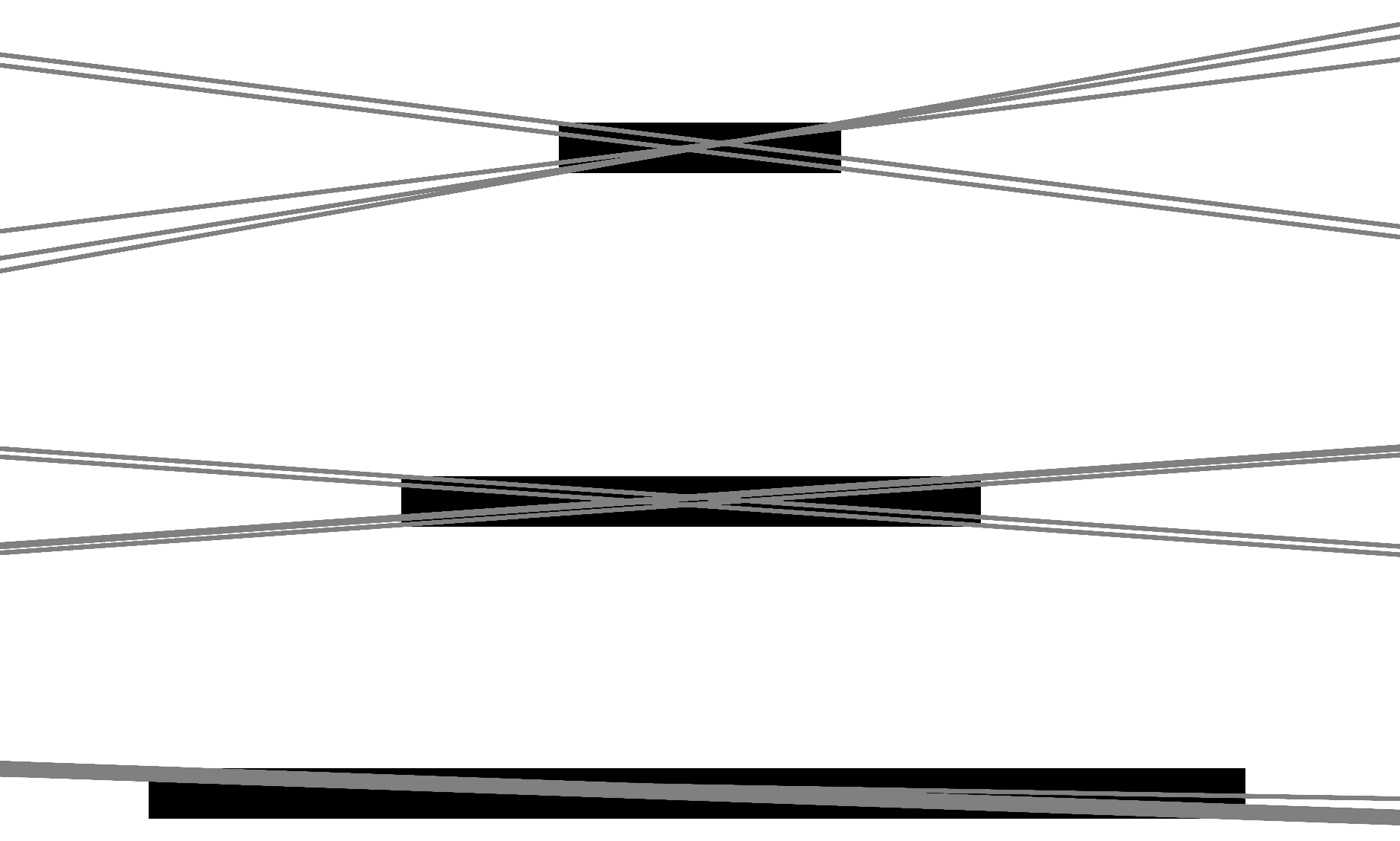} 
  \caption{Lines detected by the Hough algorithm drawn over rectangle objects representing simplified examples of linear features encountered in astronomical images. All examples use the first five maxima in the Hough space. Notice how the slope and position of detected lines changes with the rectangle shape.}
  \label{fig:thetaTresh}
\end{figure}

The ideal case for detecting linear features would be to apply the Hough transform on an image containing only edges of objects to avoid ambiguities. For example, applying the Hough line detection algorithm on an empty image with only one line being just a single pixel wide leaves no ambiguity about which set of pixels represents the line. This is unlikely in the astronomical setting, where we deal with lines of at least a dozen pixels wide. In that case we are actually trying to detected a rectangle and, as demonstrated in Figure \ref{fig:thetaTresh}, the line that encompasses the most pixels (i.e. with the most votes in Hough space) is one of the rectangle's diagonals. The number of lines fitted to a single object, or either of its diagonals, also depends on the object's width and length (see Figure \ref{fig:thetaTresh}). Our LFDA takes into account these ambiguities.

\section{Object removal}
\label{sec:objectremoval}

The purpose of this step is to remove all known objects from the image. A better object removal process yields higher detection rates and faster algorithm performance. In an ideal case where all objects in the image were registered correctly (known positions), confidently (known object shape parameters), and their removal was perfect, the result would be an empty image except for the linear features, and the detection step would be trivial. In the case of sky surveys with image differencing implemented into the data pipeline, a high quality object removal is a part of the survey's data product. LSST is going to be such an example, with linear features hiding within their transient alert stream of $\sim$10,000 events/visit ($\sim$10 million per night) \citep{LSSTCam}. The Tomo-e Gozen wide-field camera is another example \citep{Morii}, ideal for meteor surveys due to its ability to take 2 frames per second of a 20 deg$^2$ field-of-view by 84 chips of 2k$\times$1k CMOS sensors.

However, in cases when an automatic object detection pipeline is not perfect or image databases do not contain image differencing, LFDA has to incorporate its own object removal module. Depending on the set of images in question, object removal should be adapted appropriately based on the specifics of the image detector and data acquisition and availability of a catalogue of objects covered by the images. For example, the UCAC4 object catalog\footnote{\url{http://www.usno.navy.mil/USNO/astrometry/optical-IR-prod/ucac}} would contain enough objects to perform this step successfully when dealing with La Sagra Sky Survey images\footnote{\url{http://www.minorplanets.org/OLS/LSSS.html}}. For image sets with a much dimmer limiting magnitude than the 16$^\text{th}$ magnitude of the UCAC4 catalog, the object detection step can be handled by running SExtractor \citep{SExtractor} prior the star removal step to build a catalogue of objects or by extracting a list of objects from the SDSS catalogue.

However, the problem of object removal is particularly challenging in the case of SDSS, where the list of objects visible in images is obtained from the appropriate SDSS {\it photoObj} files. {\it PhotoObj} files are FITS files that contain in their header a table of all the objects registered in that particular single frame. An object is removed only if all of the following conditions are satisfied:
\begin{itemize}
\item the objects' PSF fitted magnitudes are less than the SDSS 95\% completeness limit \citep{Stoughton02} set by the LFDA parameter \emph{filter\_caps},
\item the maximum allowed object's magnitude difference between SDSS filters is larger than \emph{maxmagdiff} by at least \emph{magcount} times, where both parameters are set by the LFDA,
\item and if the number of times the object is observed was equal to the number of times the field was observed (when the same field on the sky was imaged more than once).
\end{itemize}
These conditions are fine tuned for LFDA implementation on the SDSS image database and the detailed logic behind this choice will be explained in our future publication focused solely on SDSS.

It is difficult to generalize the choice of these parameters for cases where object removal is applied on images not from SDSS. For example, setting the bottom magnitude limit \emph{filter\_caps} assures that object removal using the SDSS object catalogue does not over-mask the image. This would happen in cases where the SDSS catalogue contains a number of objects dimmer than the limiting magnitude of analysed image. In that situation object removal would mask parts of the image not showing these objects (as they are too dim), which could jeopardize the transient linear features that might appear in the image, at these same positions.

Once the objects are registered in the image a mask is constructed for object removal. Reconstructing the exact light profile for each object is also possible, but it is computationally more demanding and, as shown later in Appendix \ref{app:HoughEx}, has proved itself unnecessary. We found that masking objects with a simple square is good enough, with the square size determined by scaling the measured SDSS petrosian radius containing 90\% of the flux with the image pixel scale in arcseconds/pixel (\emph{pixscale} parameter). If the determined square is larger than the maximum allowed size (\emph{maxxy} parameter), only a default sized square is drawn whose size is set by the \emph{defaultxy} parameter. As shown in Appendix \ref{app:HoughEx}, this masking simplification has a negligible impact on the detection rates in our design of LFDA.

\begin{figure*}
\begin{subfigure}{.49\textwidth}
  \centering
  \includegraphics[width=0.98\linewidth]{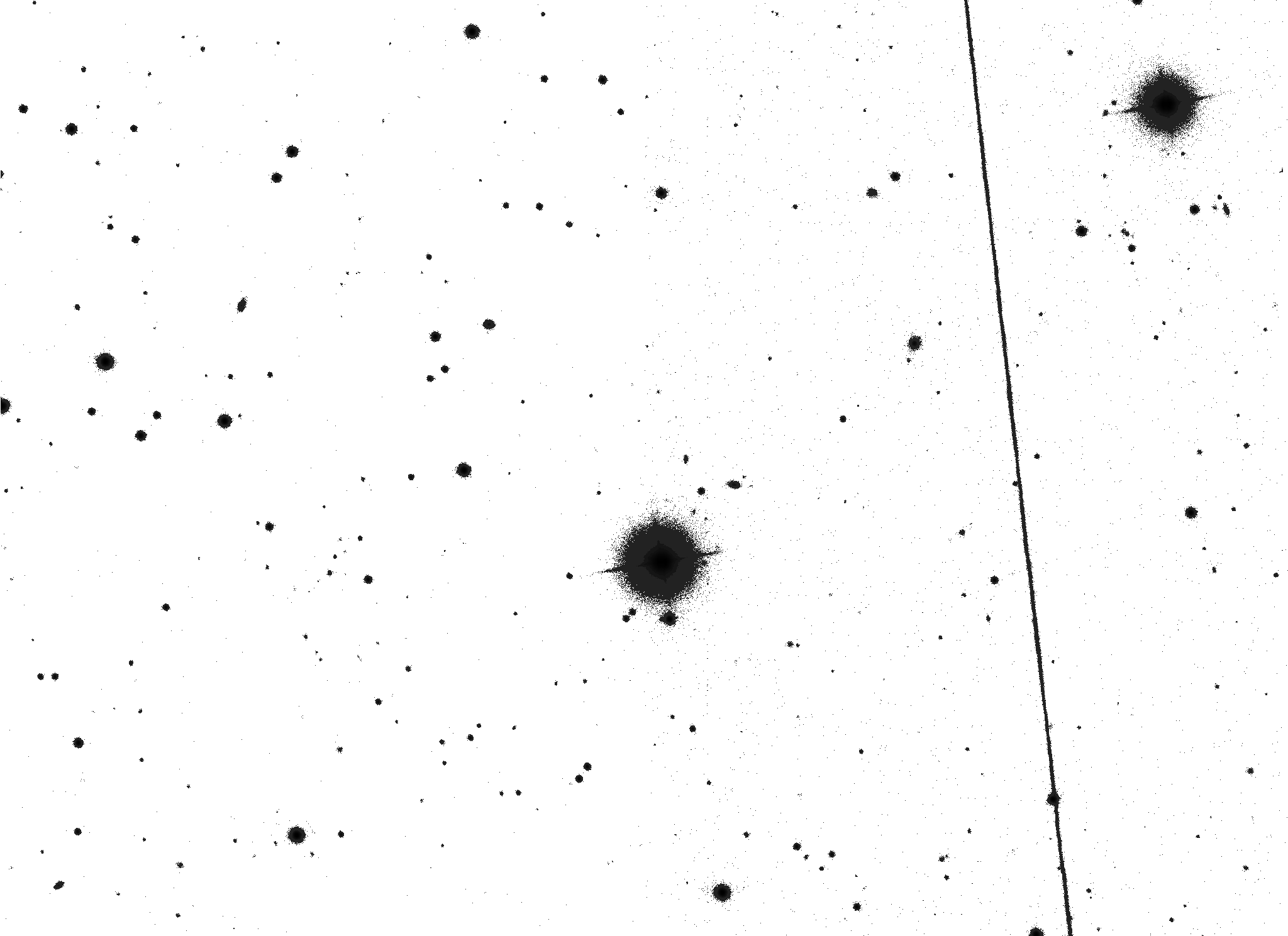}
  \caption{The original image containing a very bright trail (additional brightness adjustments need to be applied to make all the objects visible).\\}
  \label{fig:removeobj1}
\end{subfigure}
\begin{subfigure}{.49\textwidth}
  \centering
  \includegraphics[width=.98\linewidth]{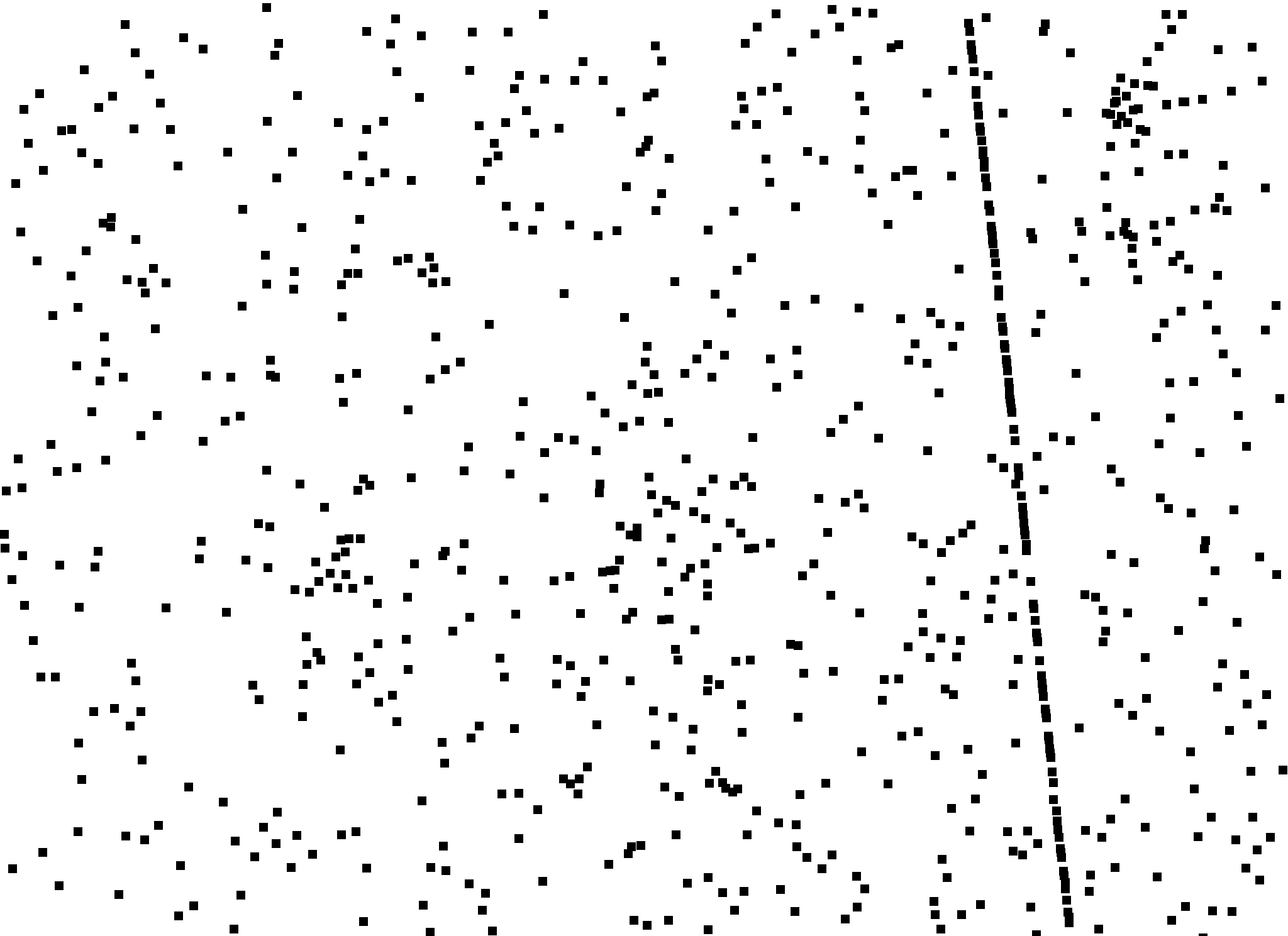}
  \caption{Mask constructed from all the objects contained in the {\it photoObj} FITS file header. Notice how the linear feature was misclassified as a large number of individual objects.}
  \label{fig:removeobj3}
\end{subfigure}
\begin{subfigure}{.49\textwidth}
  \centering
  \includegraphics[width=.98\linewidth]{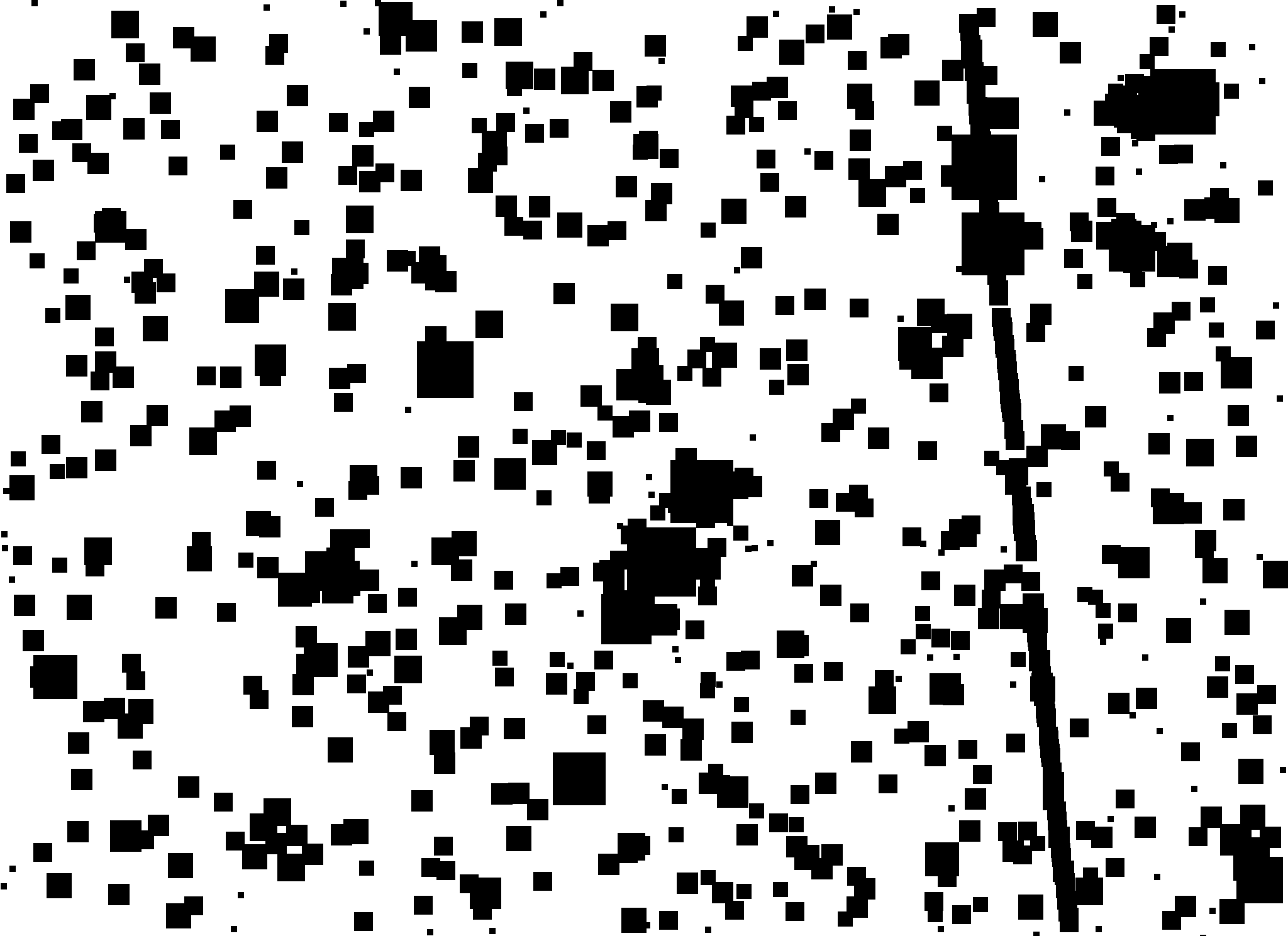}
  \caption{Applying varying square sizes to accommodate for the size of objects shows how parts of the linear feature have been detected as galaxies or stars. If this mask were used, the linear feature would be completely destroyed.}
  \label{fig:removeobj4}
\end{subfigure}
\begin{subfigure}{.49\textwidth}
  \centering
  \includegraphics[width=.98\linewidth]{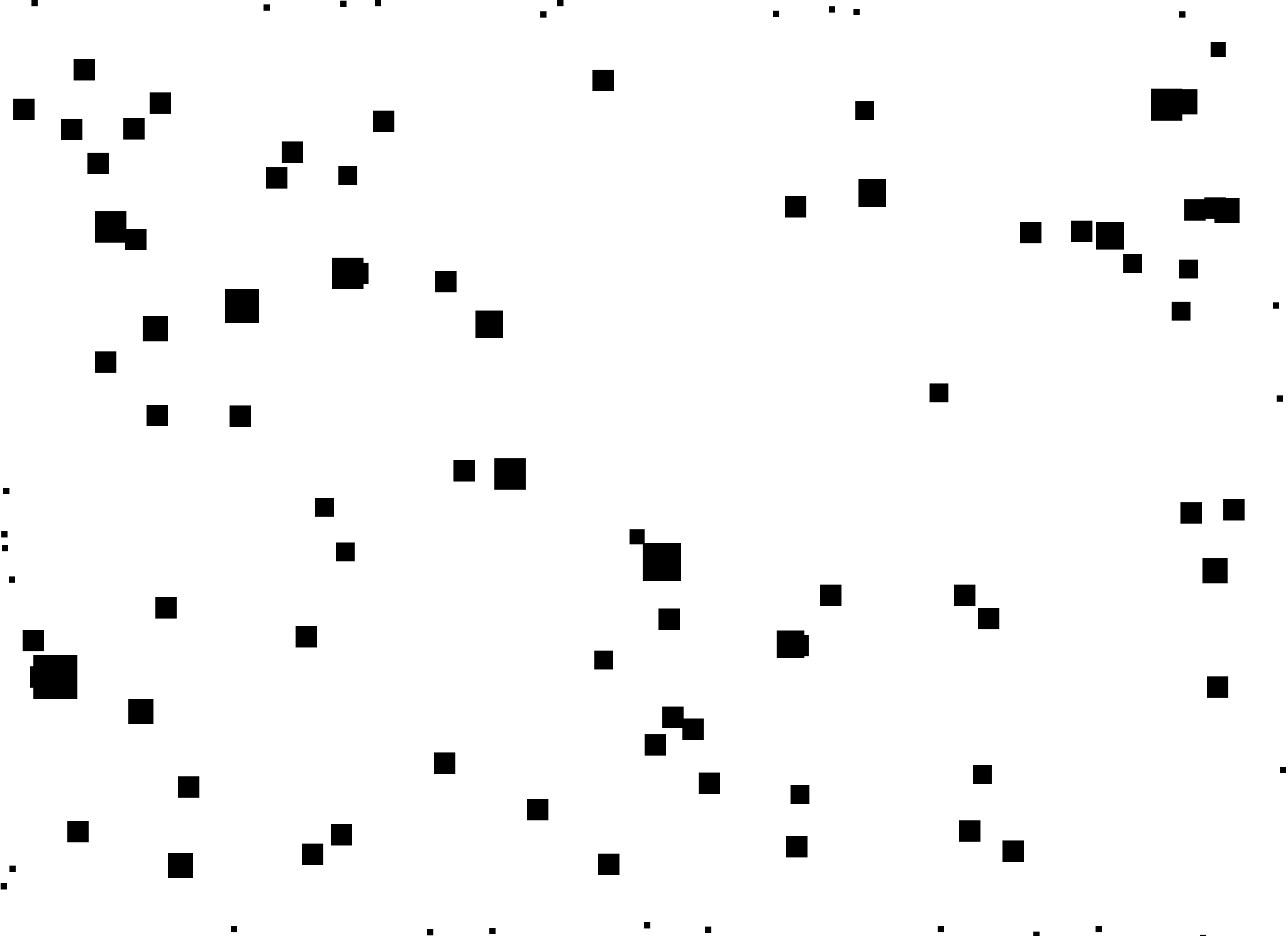}
  \caption{Imposing conditions for magnitude measurements in different SDSS filters reduces the number of objects that are masked.\\}
  \label{fig:removeobj5}
\end{subfigure}
\begin{subfigure}{.49\textwidth}
  \centering
  \includegraphics[width=.98\linewidth]{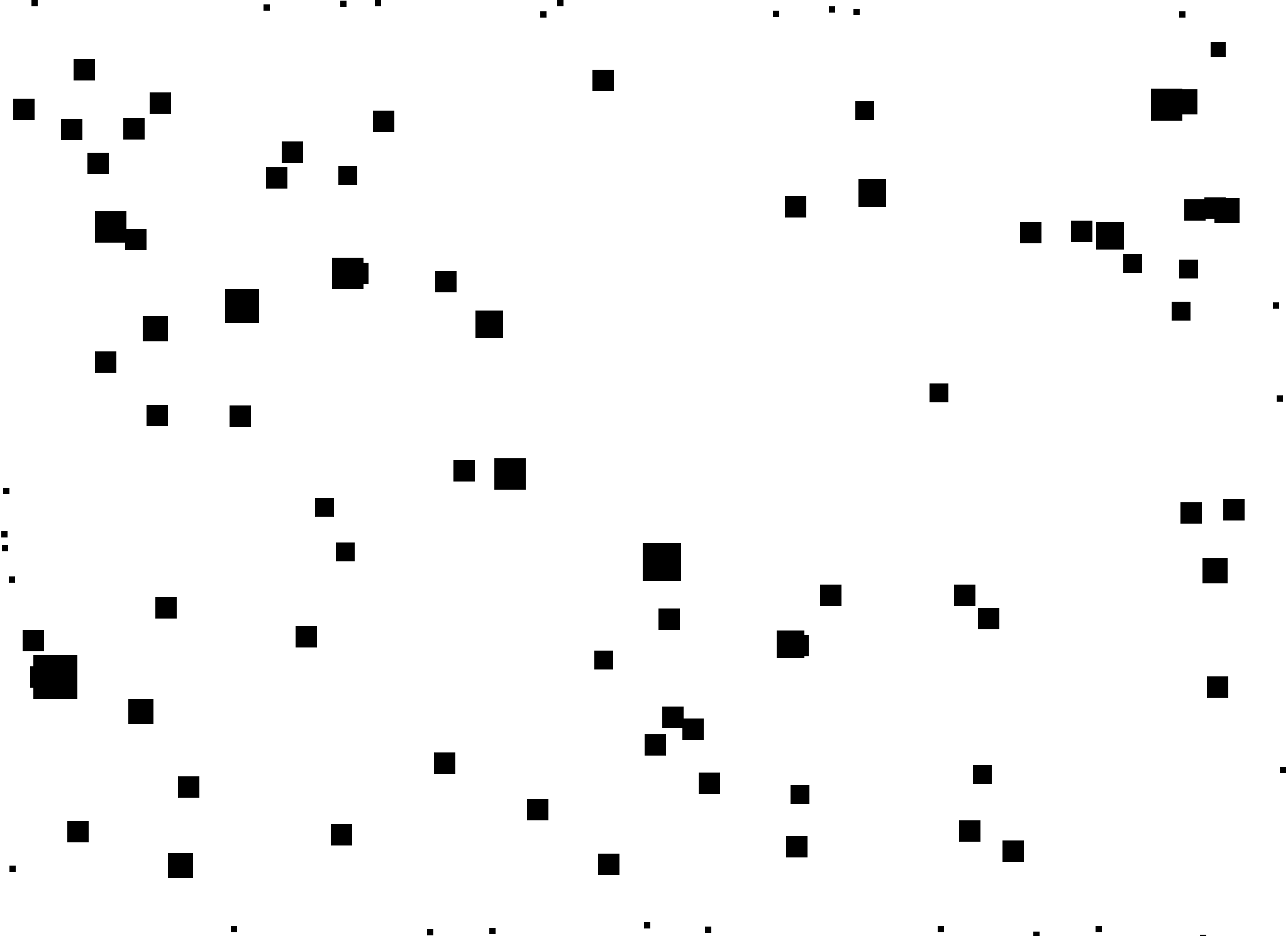}
  \caption{Equating the number of times an object and the field are observed eliminates the remaining fictitious objects.}
  \label{fig:removeobj6}
\end{subfigure}
\begin{subfigure}{.49\textwidth}
  \centering
  \includegraphics[width=.98\linewidth]{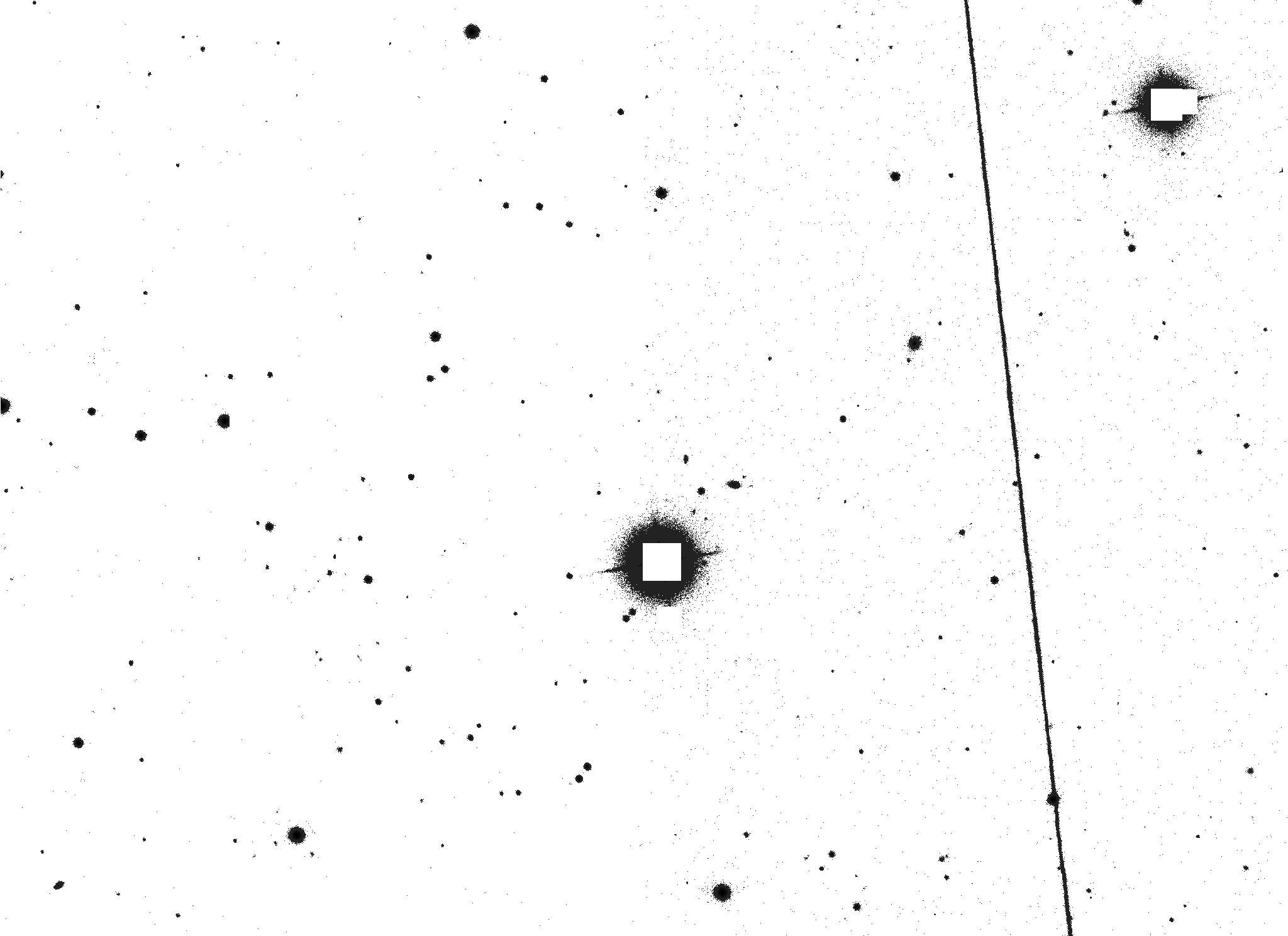}
  \caption{The resulting image once the mask is overlaid on the original image.\\.}
  \label{fig:removeobj2}
\end{subfigure}
\caption{An example of mask creation and subtraction on a particularly bright linear feature in the SDSS {\it frame-i-002888-1-0139.fits} image. The image itself has undergone slight processing to increase the brightness of the objects in the image, which otherwise would not be visible. }
\label{fig:removeobj}
\end{figure*}

This entire procedure is demonstrated in Figure \ref{fig:removeobj}. For the SDSS images these steps have proved themselves capable of removing the maximum amount of objects without harming the linear features. The features can be harmed either by accidentally covering them with oversized masks from nearby objects or by interpreting the linear feature itself as an object (see Figures \ref{fig:removeobj3} and \ref{fig:removeobj4}).

In general, where star catalogues are used, such as SDSS or UCAC4, there is no need to check for the correctness of the object detection as they have already been vetted by the survey authors. Also, the mask dimensions could be approximately determined directly from the apparent magnitudes of objects. However, if SExtractor is used for detecting objects then the dimensions of mask rectangles can be determined from the petrosian flux, but the correctness of the detection is not guaranteed and it should be carefully monitored.

\section{Detecting bright linear features}
\label{sec:DetectBright}

Bright linear features require less processing than dim features, which makes them simpler and faster to detect. In fact, excessive processing of bright features can harm the possibility of their detection. Substantially increasing the brightness and contrast of an image exposes a lot of noise and remnants of incomplete object removal. Such spurious objects would decrease the confidence in detection of linear features because they can obscure the linear feature or mask the maximum in Hough space. Detecting bright linear features is therefore primarily oriented towards detection of linear features that require minimal brightness and contrast enhancements.

All the steps in this process are listed in Figure \ref{fig:DetailedFlowChart}. It starts with reduction of the image depth, where the FITS images, usually a 32-bit float, are reduced into 8-bit integer images, with a floor function used for rounding to a whole number. New pixel values are recalculated as
\begin{equation}\label{eq:Intensities}
  I_{new}(x,y) = \min\left(\max(I_{old}(x,y), 0), 255\right)
\end{equation}
The loss of bit depth information during the conversion is not particularly concerning because the bit depth data is mostly needed for photometry information. As long as the objects' intensities are bright enough not to be rounded to zero their shape will not change drastically.

In essence, all pixels with values less than 1 are rounded to 0 and all pixels with brightness larger than 255 are rounded to 255. Pixels with brightness in the 0-255 range are left untouched. The image is then subjected to histogram equalization, followed by the dilation procedure to increase the size of all remaining objects. The size of dilation kernel is controlled by the \emph{dilateKernel} parameter found in the {\it Detecttrails} module. We refer to this converted, equalized and dilated image as the "processed image". It will be used later for line-detection by the Hough line-detection procedure.

The processed image is manipulated further to obtain an image with sufficiently elongated objects replaced by rectangles. The processed image is first exposed to the Canny edge-detection algorithm, with the hysteresis thresholding parameters set to the maximum of 255 and the minimum of 0. Effectively, this step returns all existing edges found as if hysteresis thresholding was skipped. The purpose of this edge detection is to clearly identify all remaining objects on the image in such a way that there would be no ambiguity about the position of objects' borders. This is important because the next step is extracting the shapes of objects from their contours.

The contour function's \emph{mode} and \emph{method} options are controlled by the LFDA's \emph{contoursMode} and \emph{contoursMethod} parameters. \emph{ContoursMode} is set to the RETR\_LIST flag, which means that all found contours are returned without establishing any hierarchical relationships. \emph{ContoursMethod} is set to the CHAIN\_APROX\_NONE flag, which means that all contour points are returned with no truncation or approximation. It is not recommended to change these values as the other available parameter flags can break the LFDA algorithm. The contours are then fed into the procedure for finding the minimum-area rectangles. However, the minimum area rectangles are fitted to the contours only if
\begin{itemize}
\item the shorter rectangle side is longer than the \emph{minAreaRectMinLen} parameter,
\item and the ratio of longer to shorter rectangle side is larger than \emph{lwTresh}.
\end{itemize}
The first condition excludes all remaining contours that are too small to belong to a valid linear feature. The second condition filters out all minimum area rectangles not elongated enough to be considered as a linear feature. The rectangles that pass those criteria are drawn on a new empty image. If there are no rectangles found, either because no edges or no contours were found, this detection step terminates as if the image was rejected.

Lines are derived by the Hough line detection procedure from the image with rectangles and from the processed image. This creates two sets of lines - one for the processed image and one for the image with rectangles.
The number of lines in a set is  the number of the first \emph{nLinesInSet} strongest lines in Hough space. The sets are compared to each other and the lines are compared within their sets.  When comparing lines within the same set, the spread is determined as the difference between the maximum and minimum $\theta$ coordinates. If that difference is larger than \emph{thetaTresh} then the detection is rejected. As shown in Figure \ref{fig:thetaTresh}, wider and shorter the linear features are expected to have larger spreads. We noticed that this step helps in differentiating between the linear features of interest and the star diffraction spikes. The diffraction spikes, being generally shorter than interesting linear features, have wider spreads. Setting \emph{thetaTresh} parameter to a smaller value favors the detection of longer linear features and rejects false positives generated by diffraction spikes, albeit at a risk of rejecting valid linear features. Unfortunately, it is not always possible to cleanly differentiate between diffraction spikes and linear features of interest.

Sets of lines are compared to each other in a similar way. First, all lines in a set are averaged to obtain the average $(r, \theta)$ for the set. After that the averages of two sets are compared such that the difference between average $\theta$ must not be greater than \emph{lineSetTresh} or the detection will be rejected. Since $\theta$ describes only the slope, it is also possible that the two sets of lines are nearly parallel to each other, but at different places in the image. Therefore, if the difference between average $r$ coordinates is larger than \emph{dro} then the detection is rejected. We see that working with two line sets enables us to double check that a detected linear feature is not just a spurious event.

\section{Detecting dim linear features}
\label{sec:DetectDim}

If no bright linear features were found then faint features on the image are boosted in hopes one of them is a trail. The process is similar to the detection of bright lines, except that now the first step transforms all pixels with values smaller than \emph{minFlux} to zero, while the remaining pixels have an \emph{addFlux} value added to them. Only then the image is converted from a 32-bit float image into an 8 bit image using equation (\ref{eq:Intensities}) and its contrast is enhanced by histogram equalization. Since the pixel values have been altered before the conversion and histogram equalization, a larger number of pixels remains visible in the final image.

Thus, these additional operations enhance the noise which now has to be removed as well. Increasing the \emph{minFlux} parameter would achieve this, but at the risk of destroying existing dim features. Another approach, as demonstrated in section \ref{subsec:ErosionDilatation}, is to apply the opening operator. However, we use an operation that is not a true opening as the dimensions of the erosion and dilation kernel are not equal. The problem with certain dim linear features is that they are "transparent" - so dim that they are barely distinguishable from the background. This creates a problem for erosion because pixels within transparent objects can be eroded. When that happens dilation has no effect because there is no object center to expand from. To combat these effects it is necessary to use a small erosion kernel and a relatively large dilatation kernel in order to try to preserve and restore as much of the original linear feature as possible. The dimension of the erosion kernel is controlled by the \emph{erosionKernel} parameter, while \emph{dilatationKernel} controls the size of dilation kernel.

The remaining detection steps are the same as when detecting bright linear features: detection of edges by Canny, locating contours among the edges and then a reconstruction of the image via minimum-area-rectangles approach. Lines are fitted to both the reconstructed image and the processed original image and their spread is tested individually. The two line sets are then compared to each other in the same way as in the detection of bright lines. The parameters involved are named the same as before, but do not necessarily have the same values.

\section{LFDA processing examples}
\label{sec:Examples}

\begin{figure*}
\begin{subfigure}{.49\textwidth}
  \centering
  \includegraphics[width=0.98\linewidth]{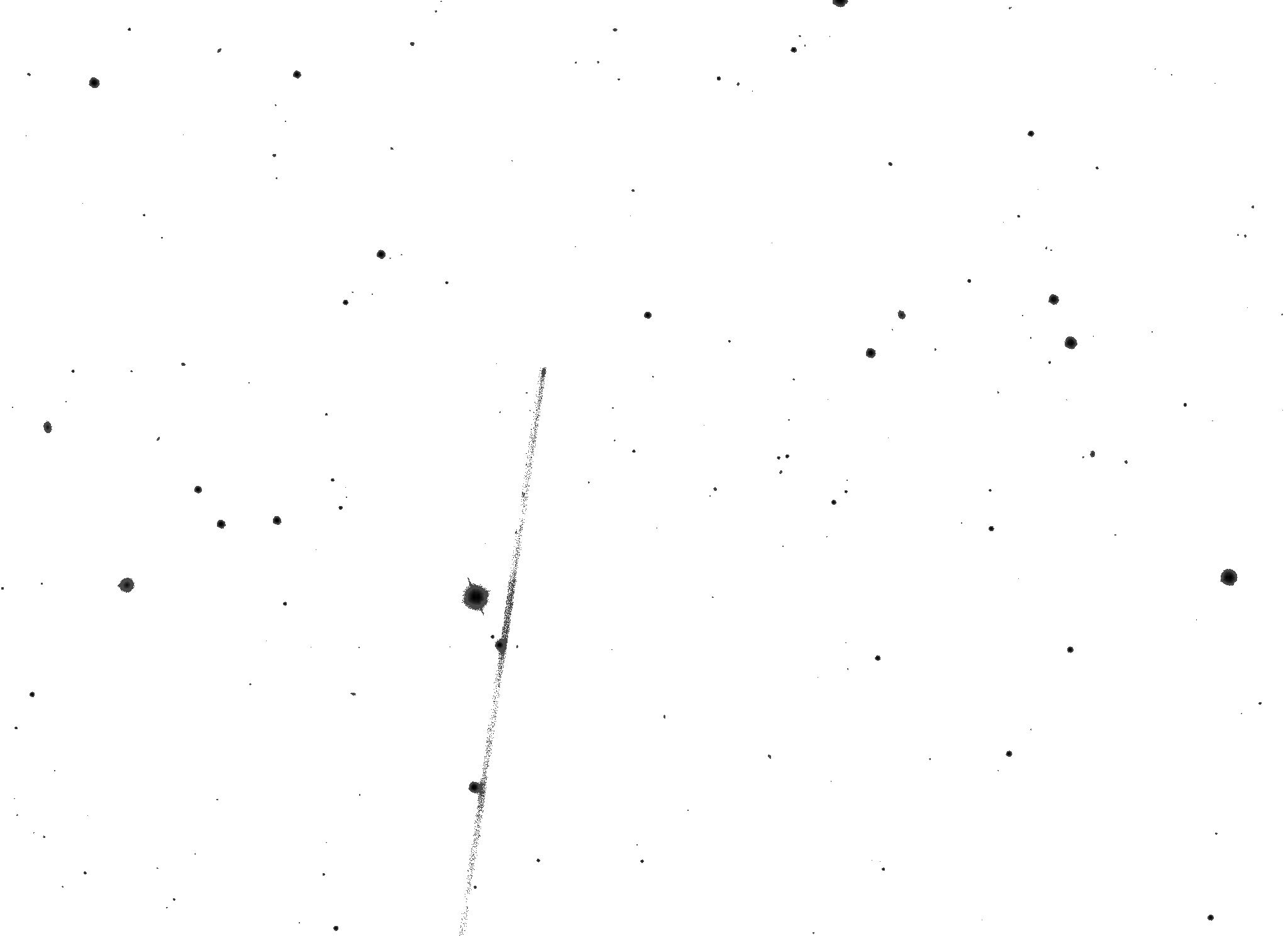}
  \caption{The original image with a relatively bright trail. The image has undergone brightness and contrast enhancements.}
  \label{fig:0origBRIGHT}
\end{subfigure}
\begin{subfigure}{.49\textwidth}
  \centering
  \includegraphics[width=.98\linewidth]{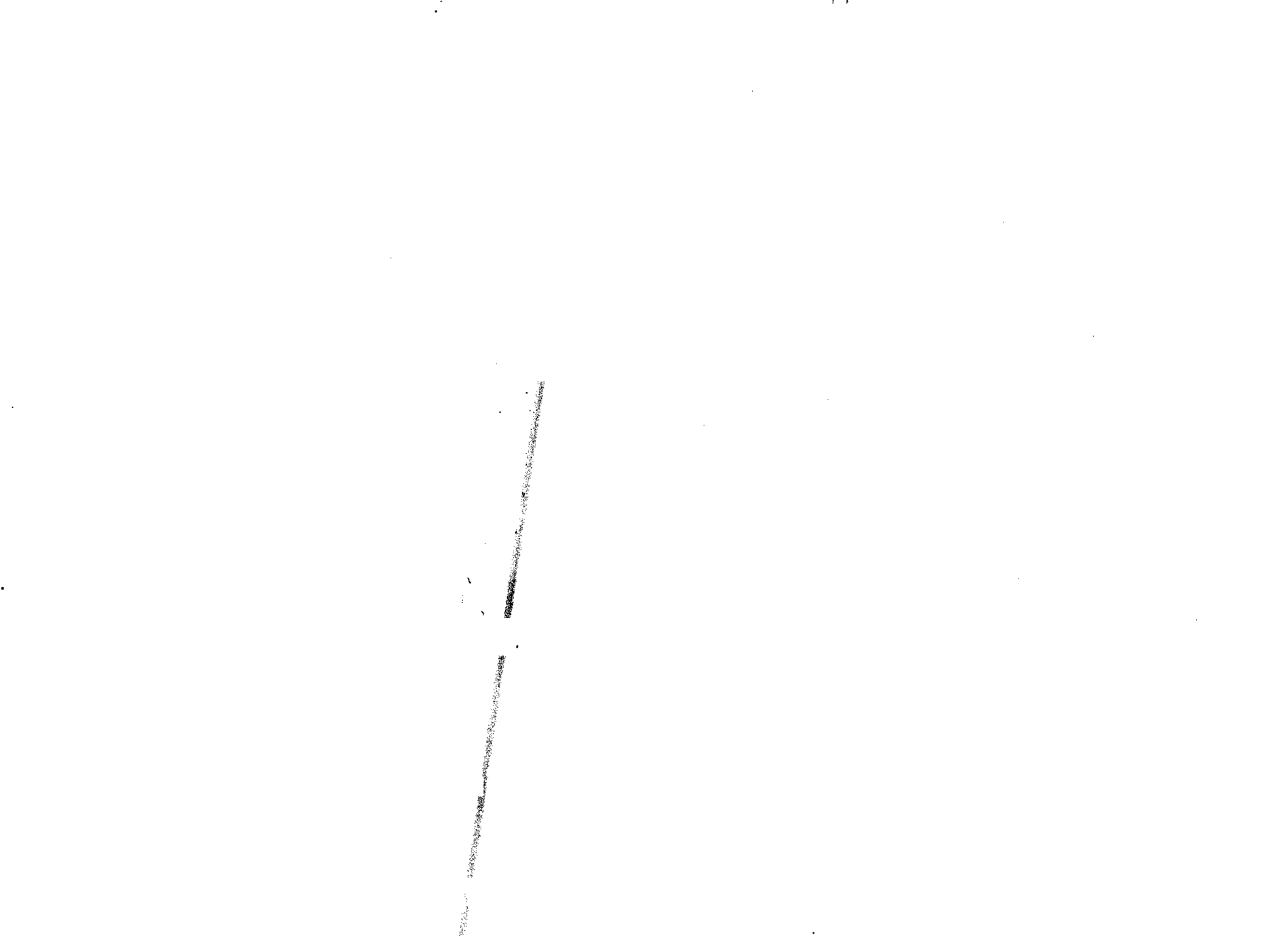}
  \caption{Objects are masked, the image is converted to 8-bit image and then histogram equalized.}
  \label{fig:1equBRIGHT}
\end{subfigure}
\begin{subfigure}{.49\textwidth}
  \centering
  \includegraphics[width=.98\linewidth]{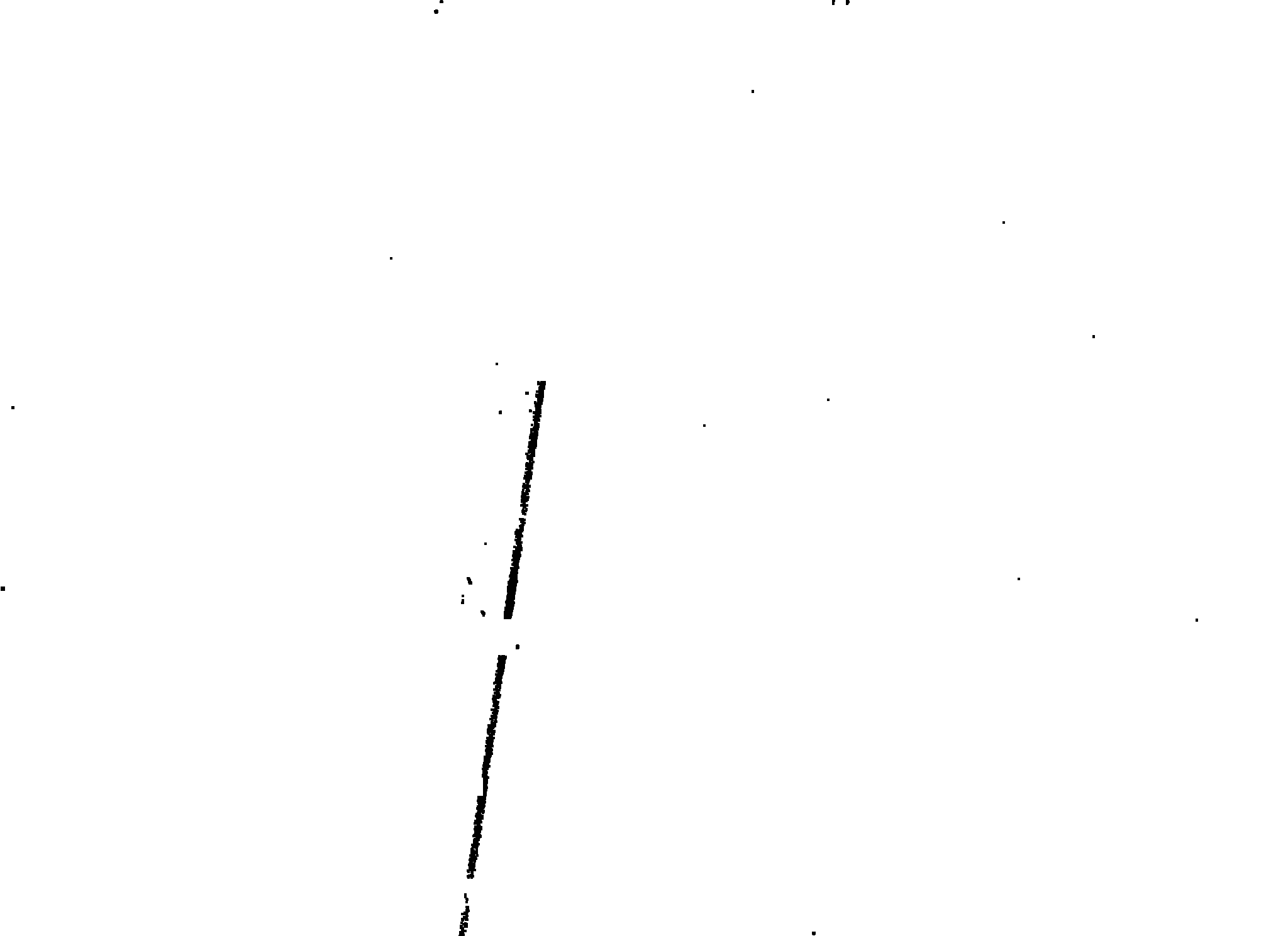}
  \caption{A dilation operator is applied using a 4x4 kernel in order to enhance the objects and fill-in any holes due to line transparency after conversion.}
  \label{fig:2dilateBRIGHT}
\end{subfigure}
\begin{subfigure}{.49\textwidth}
  \centering
  \includegraphics[width=.98\linewidth]{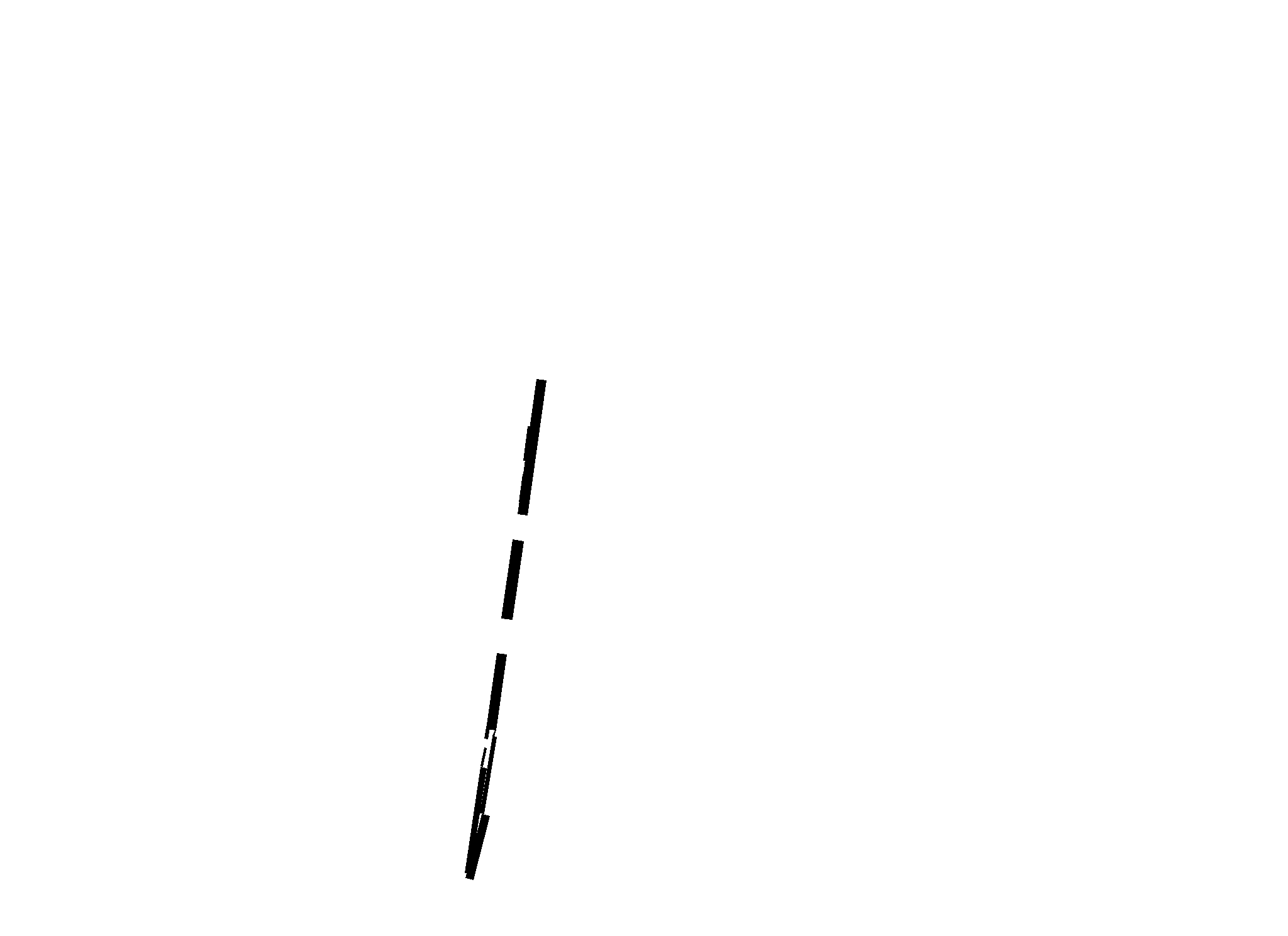}
  \caption{After the edges and contours have been found the processed image from Figure \ref{fig:2dilateBRIGHT} is reconstructed by drawing only the elongated minimum area rectangles.}
  \label{fig:3contoursBRIGHT}
\end{subfigure}
\begin{subfigure}{.49\textwidth}
  \centering
  \includegraphics[width=.98\linewidth]{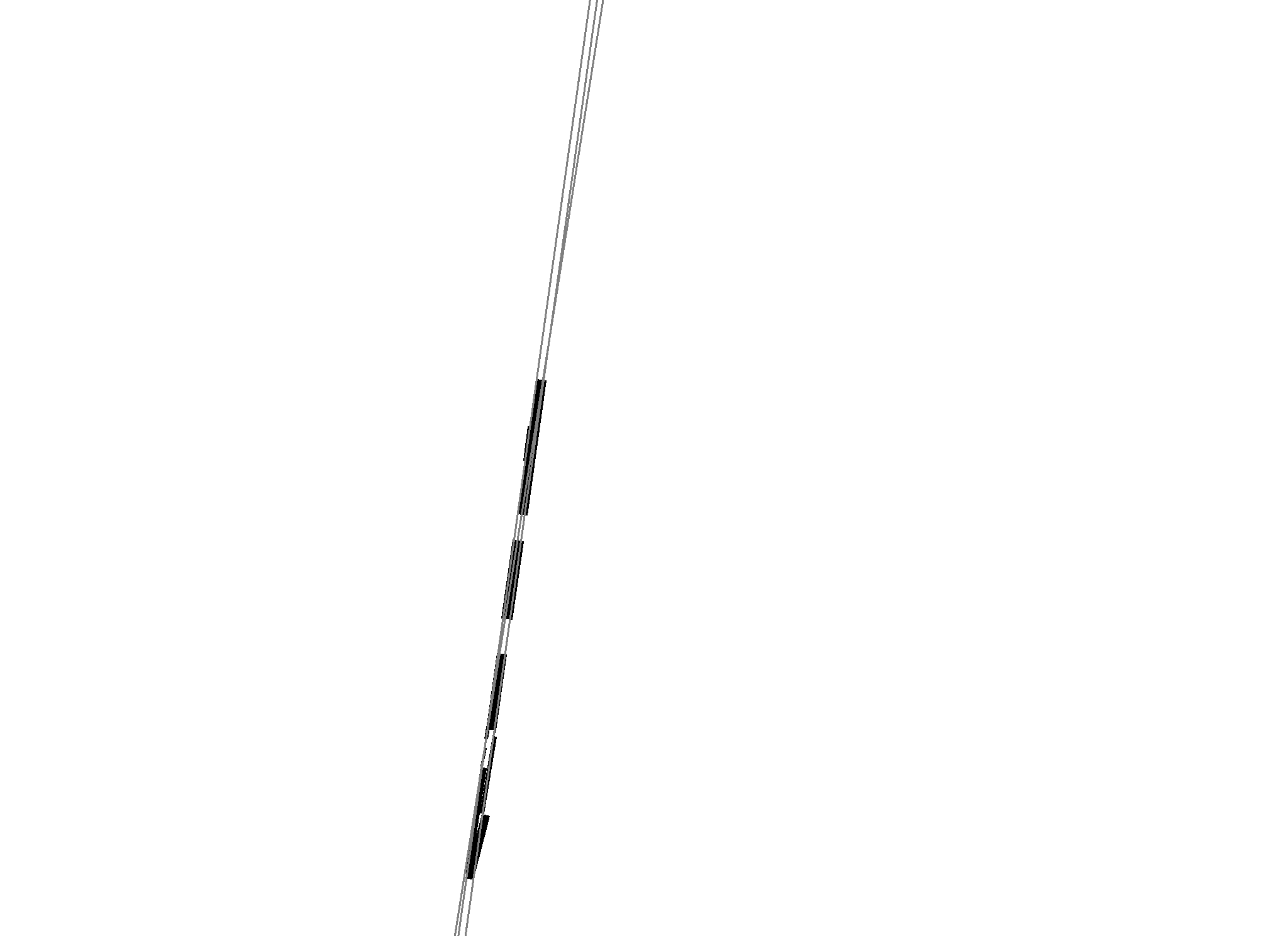}
  \caption{Lines are fitted to the reconstructed image in Figure \ref{fig:3contoursBRIGHT}.}
  \label{fig:4boxhoughBRIGHT}
\end{subfigure}
\begin{subfigure}{.49\textwidth}
  \centering
  \includegraphics[width=.98\linewidth]{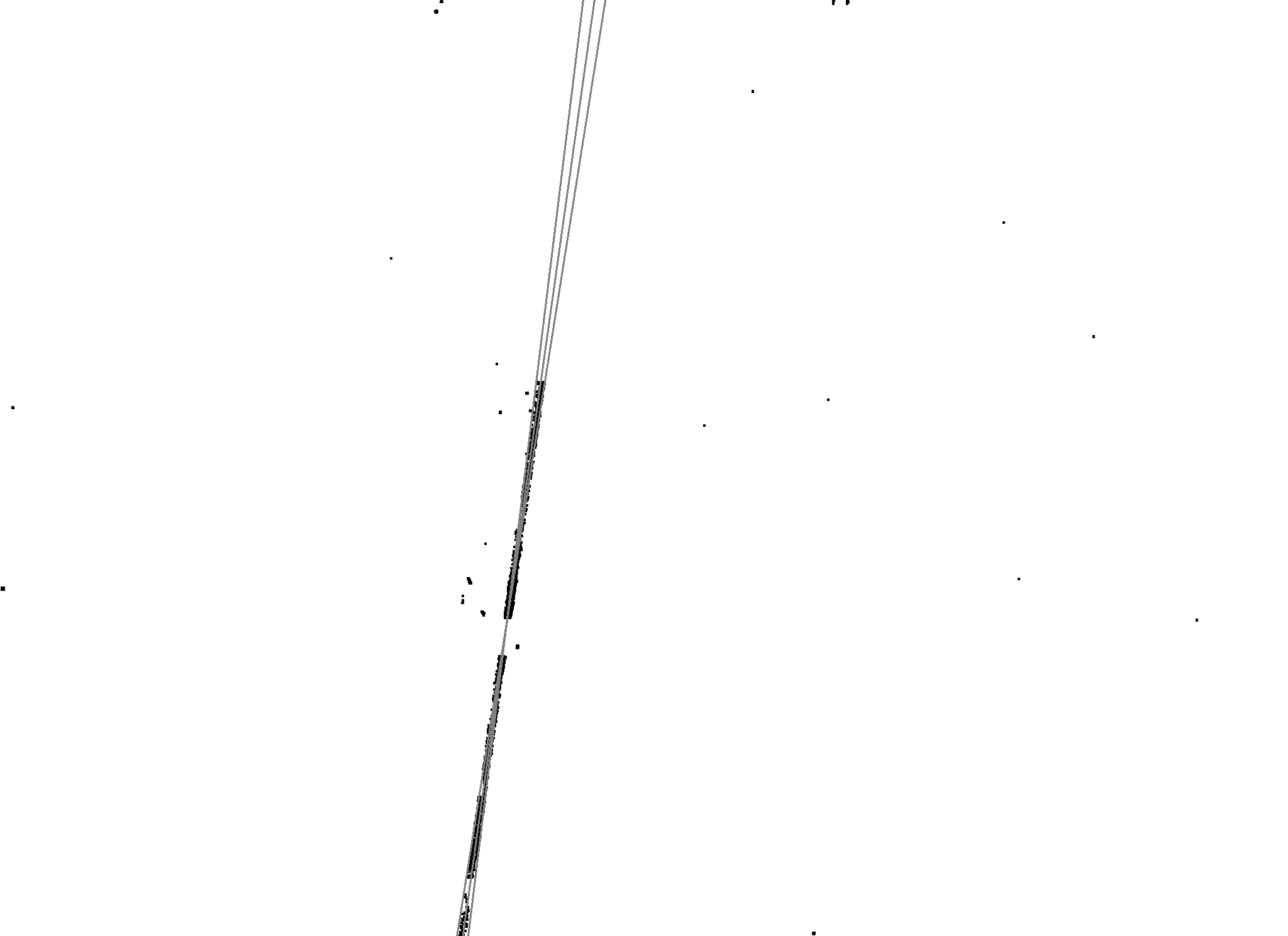}
  \caption{Lines are fitted to the processed image in Figure \ref{fig:2dilateBRIGHT}.}
  \label{fig:5equhoughBRIGHT}
\end{subfigure}
\caption{An example of how bright detection algorithm processes bright trails. The example uses the SDSS {\it frame-i-005973-3-0130} image frame. The original image in Figure \ref{fig:0origBRIGHT} had its brightness and contrast increased in the same way as in Figure \ref{fig:1equBRIGHT} in order to make the objects in the image more pronounced. After fitted lines from Figure \ref{fig:4boxhoughBRIGHT} and Figure \ref{fig:5equhoughBRIGHT} were tested and compared (see section \ref{sec:DetectBright}) the detection was accepted.}
\label{fig:brightdet}
\end{figure*}

We illustrate the LFDA processing sequence for three situations. Figure \ref{fig:brightdet} shows the steps involved in a successful detection of a bright linear feature. Without dilating the image, there is a strong possibility that no minimum area rectangles would be found due to the transparency of the linear feature. Notice also the low level of noise in the image. This example is a limiting case of a successful detection of bright features, as trails slightly dimmer or more transparent require additional steps for detecting dim features.

\begin{figure*}
\begin{subfigure}{.49\textwidth}
  \centering
  \includegraphics[width=0.98\linewidth]{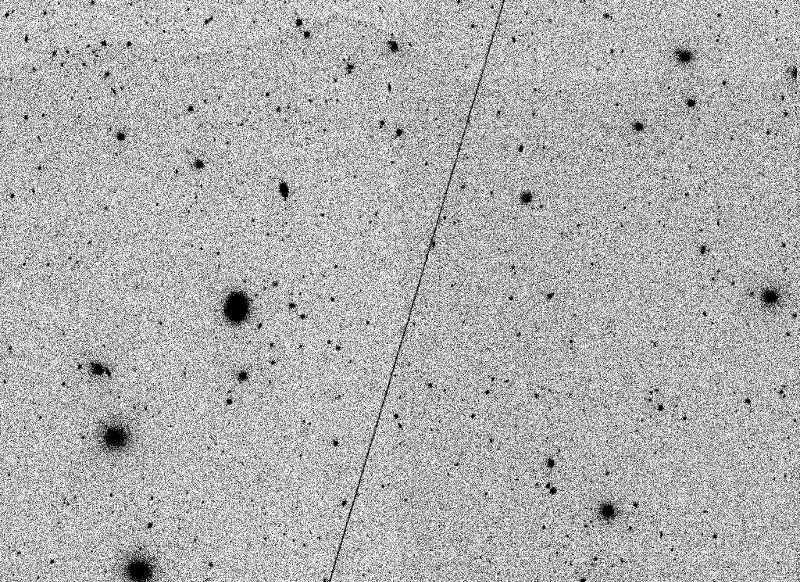}
  \caption{The original image with a dim trail. The image has undergone brightness and contrast enhancements.}
  \label{fig:0origDIM}
\end{subfigure}
\begin{subfigure}{.49\textwidth}
  \centering
  \includegraphics[width=.98\linewidth]{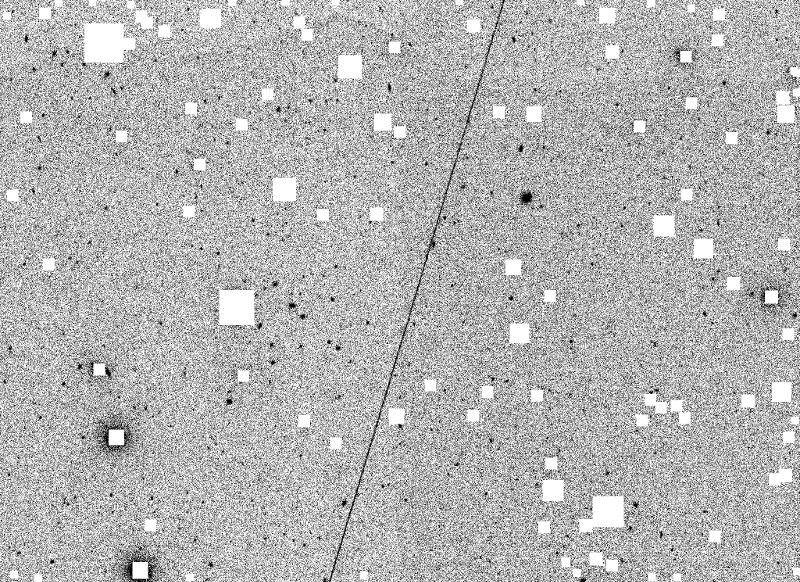}
  \caption{Objects are masked, brightness is increased, the image is converted to 8-bit image and then histogram equalized.}
  \label{fig:1equDIM}
\end{subfigure}
\begin{subfigure}{.49\textwidth}
  \centering
  \includegraphics[width=.98\linewidth]{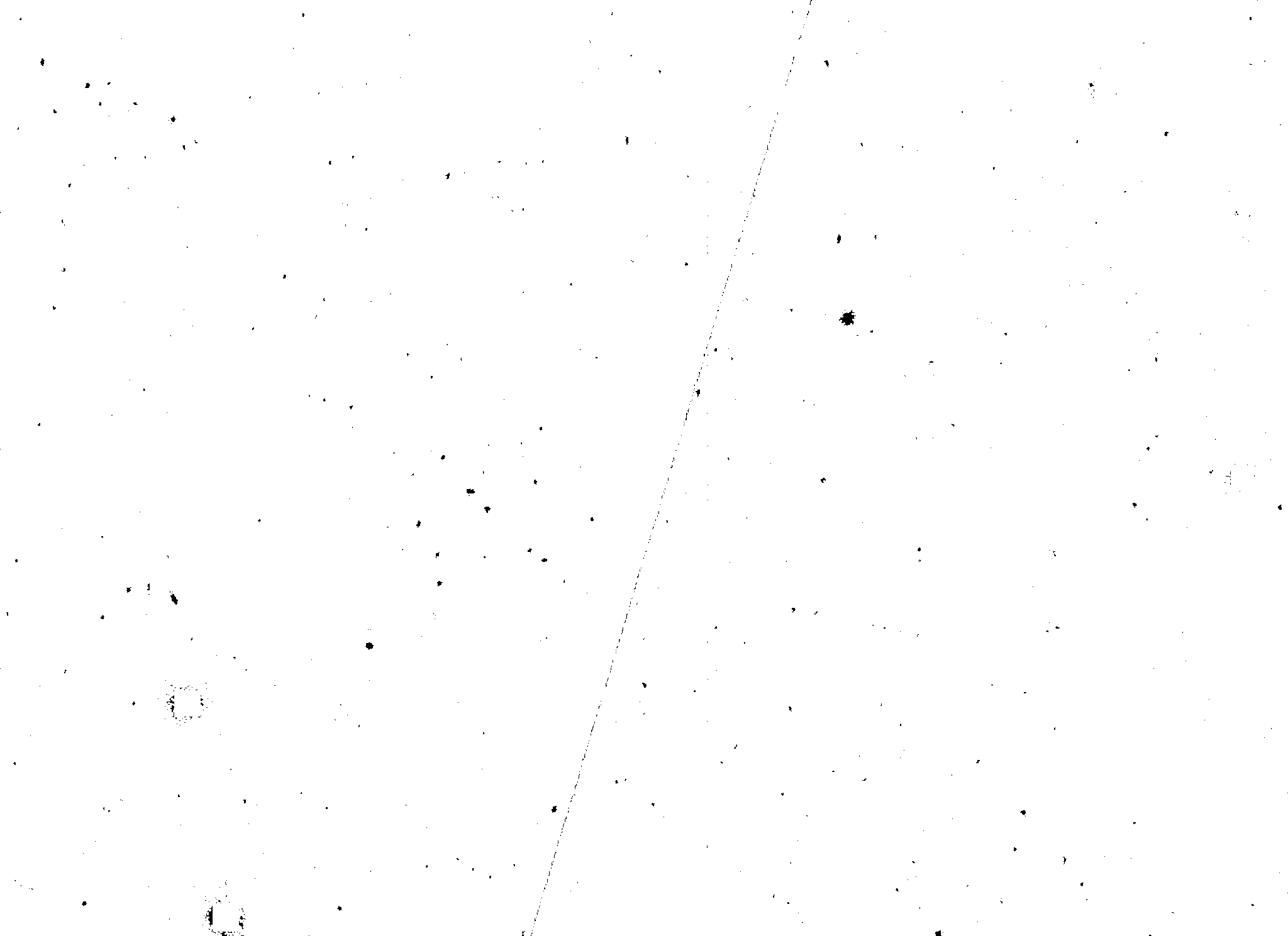}
  \caption{An erosion operator is applied using a 3x3 kernel in order to remove noise now present due to brightness and contrast enhancements.}
  \label{fig:2erodedDIM}
\end{subfigure}
\begin{subfigure}{.49\textwidth}
  \centering
  \includegraphics[width=.98\linewidth]{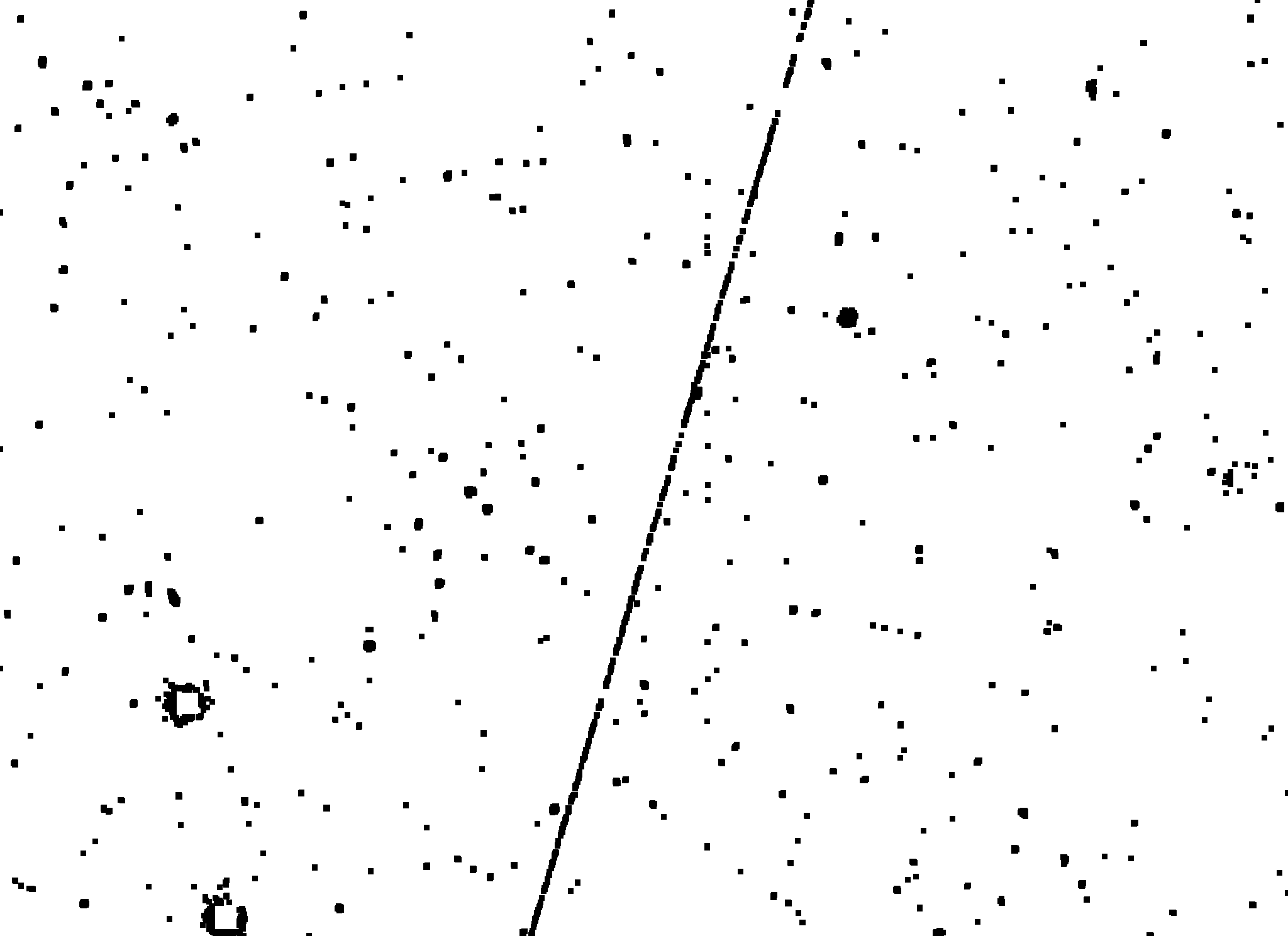}
  \caption{A dilation operator is applied using a 9x9 kernel in order to enhance the objects, reconnect them and fill-in any holes due to erosion.}
  \label{fig:3openedDIM}
\end{subfigure}
\begin{subfigure}{.49\textwidth}
  \centering
  \includegraphics[width=.98\linewidth]{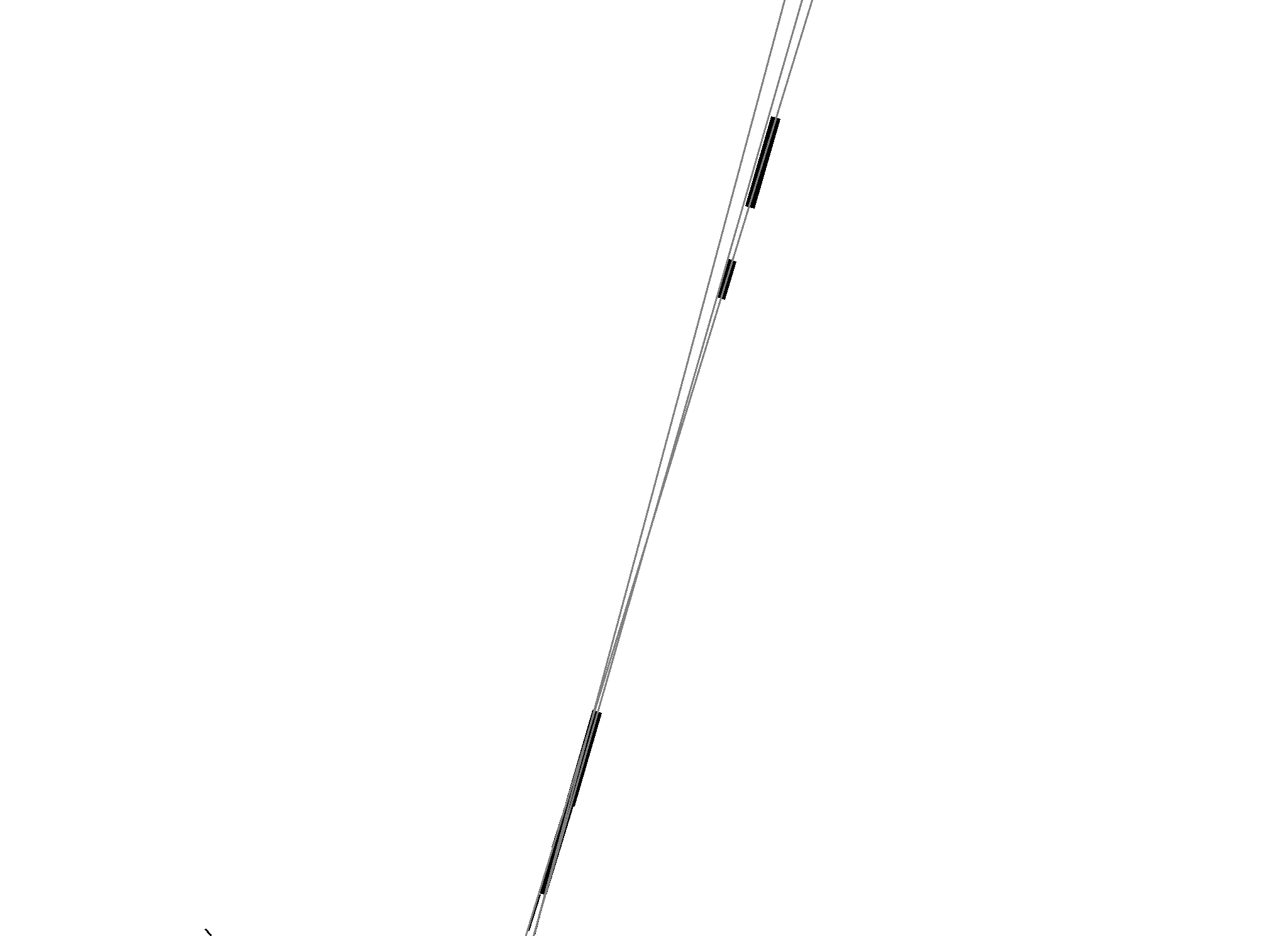}
  \caption{Contours are found among Canny edges, followed by the minimum area rectangles fit to Figure-\ref{fig:3openedDIM} and finally the lines are fitted to the rectangles.}
  \label{fig:4contoursDIM}
\end{subfigure}
\begin{subfigure}{.49\textwidth}
  \centering
  \includegraphics[width=.98\linewidth]{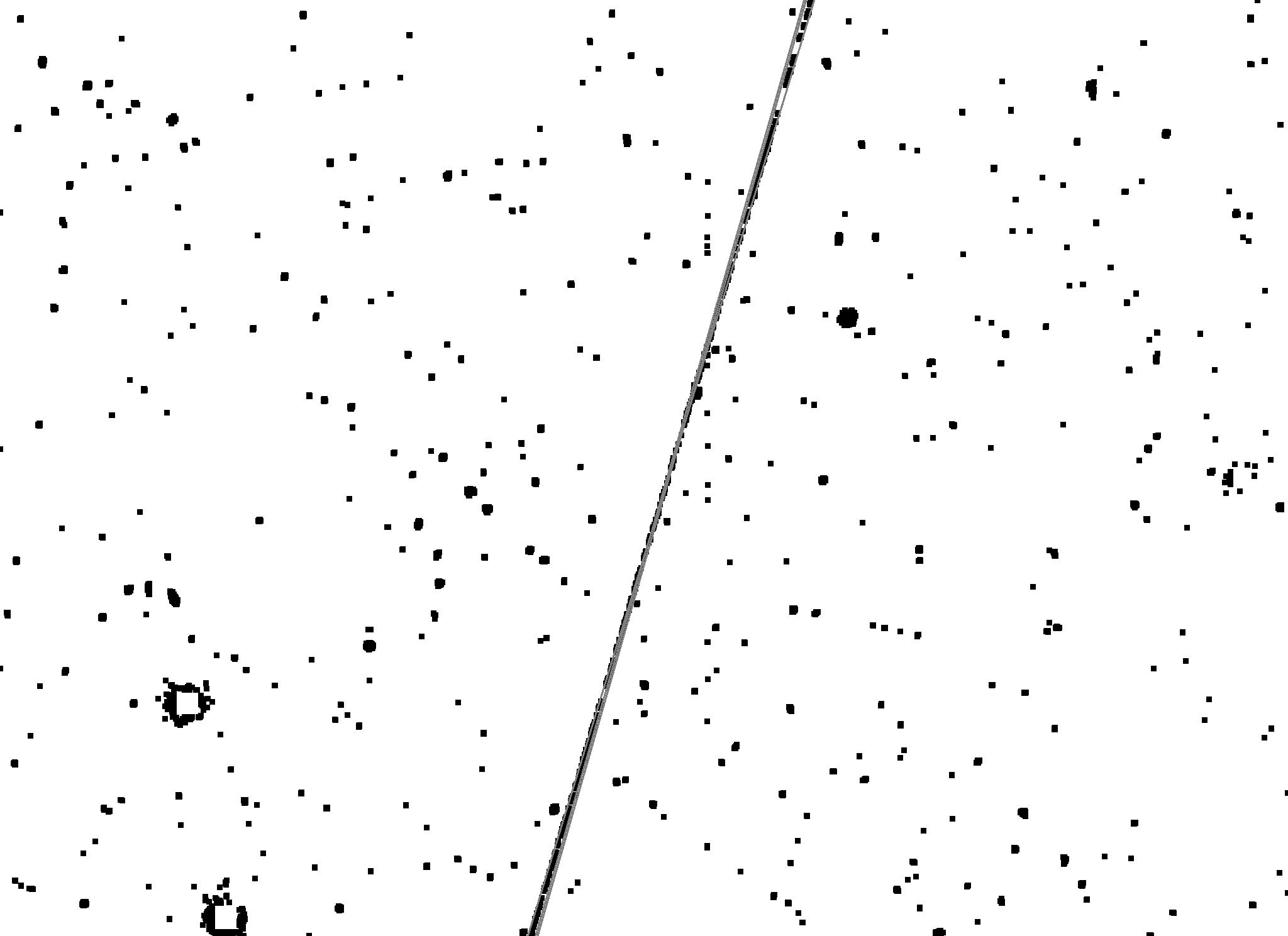}
  \caption{Lines are fitted to Figure \ref{fig:2erodedDIM} and drawn on Figure \ref{fig:3openedDIM}.}
  \label{fig:5equhoughDIM}
\end{subfigure}
\caption{An example of how dim detection algorithm processes dim trails. The examples uses the SDSS {\it frame-i-000094-1-0313} image frame. The original image in Figure \ref{fig:0origDIM} had its brightness and contrast increased in the same way as in Figure \ref{fig:1equDIM} in order to make the objects in the image more pronounced. After fitted lines from Figure \ref{fig:4contoursDIM} and Figure \ref{fig:3openedDIM} were tested and compared (see section \ref{sec:DetectDim}) the detection was accepted.}
\label{fig:dimdet}
\end{figure*}

Figure \ref{fig:dimdet} represents steps taken during a successful detection of the dim linear feature. This procedure is activated after a failed attempt to detect a bright feature. The linear feature visible in the image is very thin and not particularly transparent. This example displays the dangers of erosion step, where trails that are more transparent or thinner would not be detected because erosion destroys them. However, if the erosion step were omitted, the sheer amount of the background noise now present in the image would negatively impact the line detection and would have a drastic impact on the execution time required due to a larger number of active pixels.

\begin{figure*}
\begin{subfigure}{.49\textwidth}
  \centering
  \includegraphics[width=0.98\linewidth]{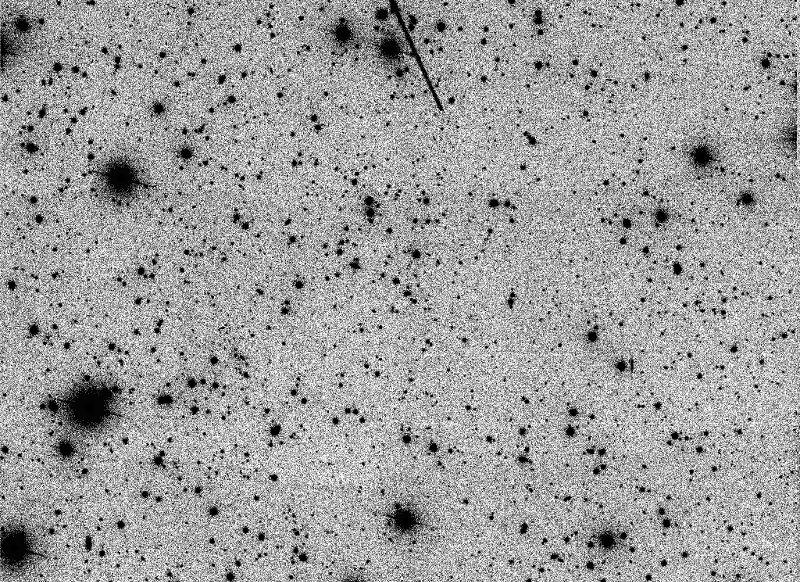}
  \caption{The original image with a dim trail. The image has undergone brightness and contrast enhancements.}
  \label{fig:0origNEG}
\end{subfigure}
\begin{subfigure}{.49\textwidth}
  \centering
  \includegraphics[width=.98\linewidth]{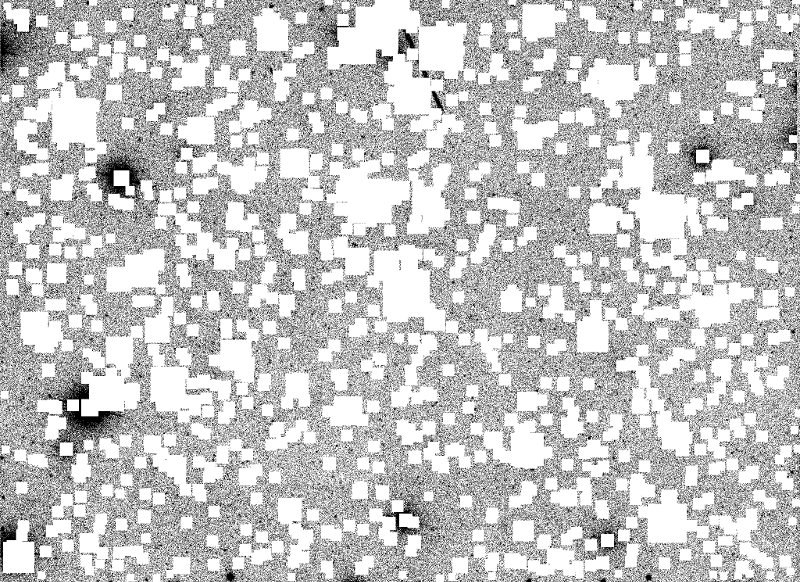}
  \caption{Objects are masked, brightness is increased, the image is converted to 8-bit image and then histogram equalized.}
  \label{fig:1equNEG}
\end{subfigure}
\begin{subfigure}{.49\textwidth}
  \centering
  \includegraphics[width=.98\linewidth]{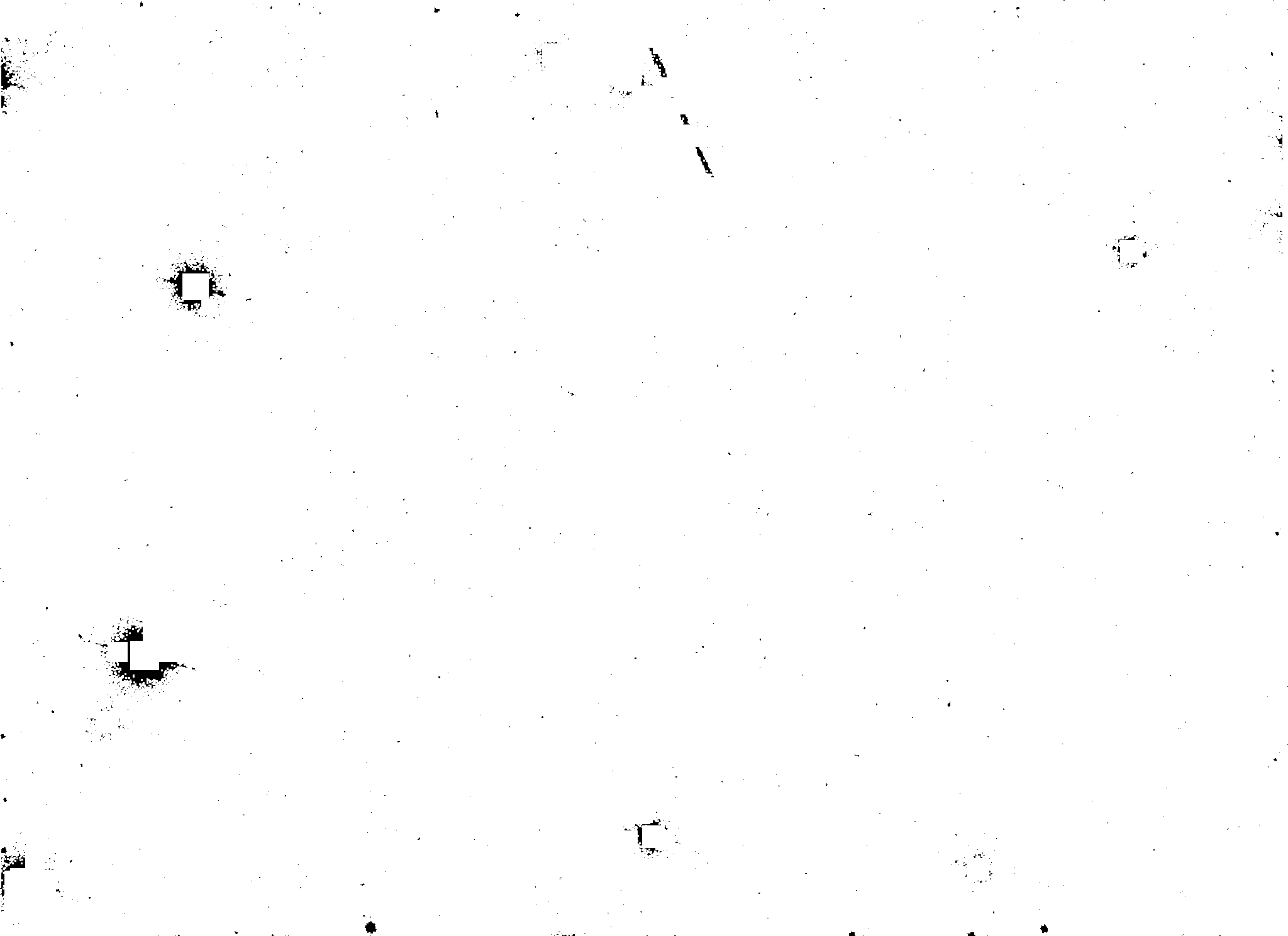}
  \caption{An erosion operator is applied using a 3x3 kernel in order to remove noise now present due to brightness and contrast enhancements.}
  \label{fig:2erodedNEG}
\end{subfigure}
\begin{subfigure}{.49\textwidth}
  \centering
  \includegraphics[width=.98\linewidth]{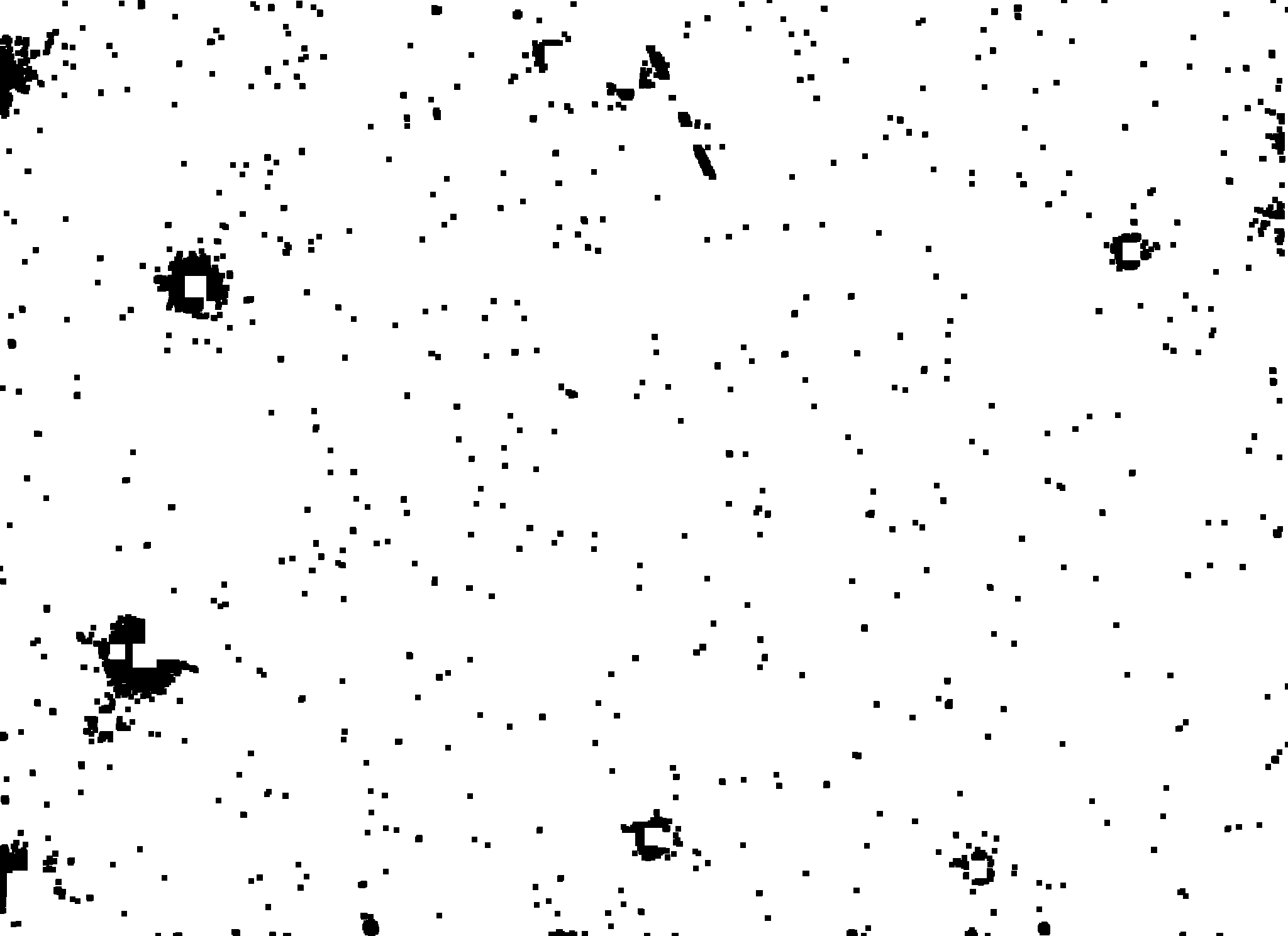}
  \caption{A dilation operator is applied using a 9x9 kernel in order to enhance the objects, reconnect them and fill-in any holes due to erosion.}
  \label{fig:3openedNEG}
\end{subfigure}
\begin{subfigure}{.49\textwidth}
  \centering
  \includegraphics[width=.98\linewidth]{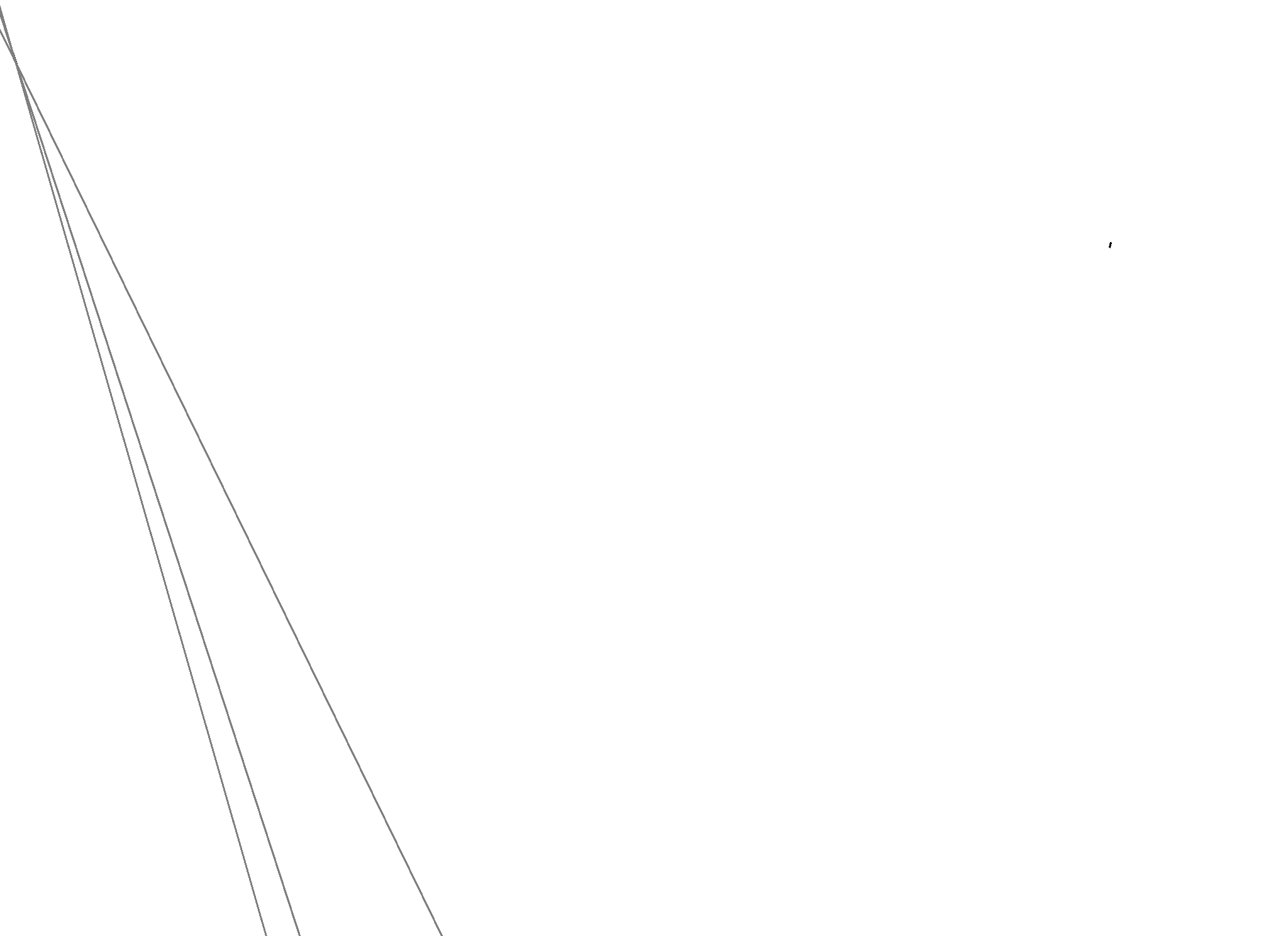}
  \caption{Contours are found among Canny edges, followed by the minimum area rectangles fit to Figure-\ref{fig:3openedNEG} and finally the lines are fitted to the rectangles.}
  \label{fig:4contoursNEG}
\end{subfigure}
\begin{subfigure}{.49\textwidth}
  \centering
  \includegraphics[width=.98\linewidth]{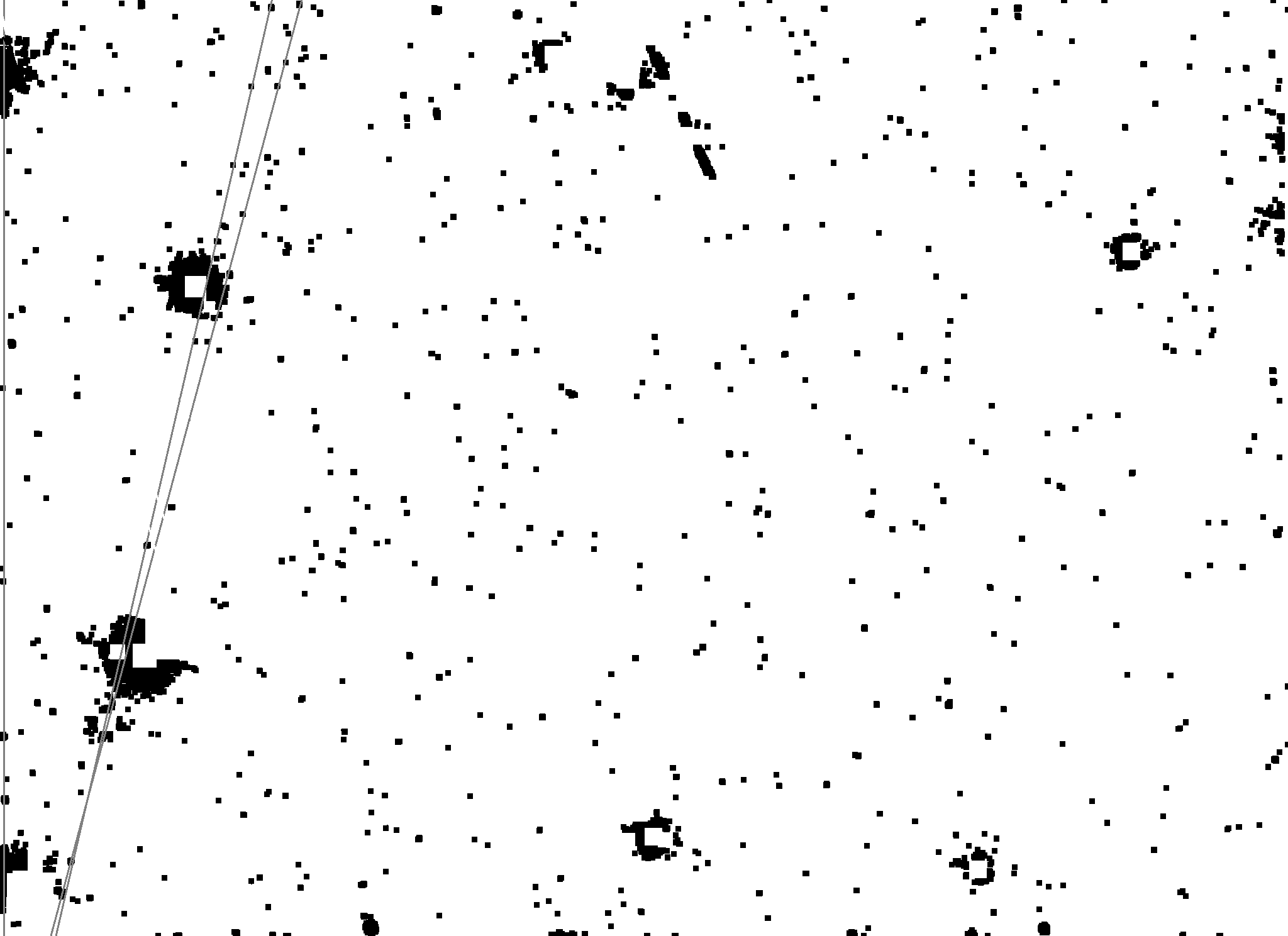}
  \caption{Lines are fitted to Figure \ref{fig:2erodedNEG} and drawn on Figure \ref{fig:3openedNEG}.}
  \label{fig:5equhoughNEG}
\end{subfigure}
\caption{An example of a false negative detection. The example uses the SDSS {frame-i-005415-1-0055} image frame. Even though a trail exists in the image it is not detected. The main reasons for failed detection is over-masking objects in a relatively dense field. Additionally, due to the shortness of the linear feature it is unclear whether \emph{thetaTresh} conditions would have been met. }
\label{fig:falseneg}
\end{figure*}

Figure \ref{fig:falseneg} represents a false-negative detection. Due to the relatively crowded field and shortness of the trail visible in the image, the detection ended up unsuccessful. It demonstrates limitations of the current version of LFDA, but also a situation where the algorithm can be further improved in future releases.

All LFDA parameter values were the same for all three figures and are shown in Table \ref{tb:detpar}. However, these detection parameters do not represent the optimal detection parameters for SDSS, which we will address in our future publication. In addition to the listed parameters, there is one more parameter \emph{debug} that controls the verbosity of execution by printing various messages to the screen as well as saving images depicting the processing steps.

\begin{table*}
\caption{Parameters that control the execution of line detection algorithm in {\it Detecttrails} module.}
\label{tb:detpar}
\begin{tabular}{p{2.8cm}|p{7cm}|p{6.1cm} }
Parameter name       & Parameter description                                                                  & Parameter value                                                            \\
\hline
                        & RemoveStars parameters                                                              &                                                                            \\
\hline
 pixscale            & Pixel scale (arcseconds/pixel)                                                         & 0.396                                                                      \\
 defaultxy           & Default mask size (pixels)                                                             & 10                                                                         \\
 maxxy               & Maximum allowed mask size (pixels)                                                     & 60                                                                         \\
 filter\_caps        & Minimum apparent magnitude under which objects are not removed, given per SDSS filter. & u:22.0, g:22.2, r:22.2, i:21.3, z:20.5                                     \\
 magcount            & Number of different filters in which magnitude difference can exceed maxmagdiff.       & 3                                                                          \\
 maxmagdiff          & Maximal magnitude difference allowed between two filters.                              & 5                                                                          \\
\hline
                     & Detect bright parameters                                                               &                                                                            \\
\hline
 dilateKernel        & Dilation kernel.                                                                       & 4x4 matrix, all elements equal to 1.                                       \\
 contoursMode        & Contours mode.                                                                         & RETR\_LIST, return all contours as a flat list.                            \\
 contoursMethod      & Contours approximation method.                                                         & CHAIN\_APROX\_NONE, a full set of points encompassing the contour is returned. \\
 minAreaRectMinLen   & Allowed minimum length of sides of a minimum area rectangle. (pixels)                  & 1                                                                          \\
 lwTresh             & Minimum allowed ratio of sides of a minimum area rectangle.                            & 5                                                                          \\
 nlinesInSet         & Number of detected lines to be declared a set.                                         & 3                                                                          \\
 thetaTresh          & Maximum allowed difference between any two line slopes in a set. (radians)             & 0.15                                                                      \\
 lineSetTresh        & Maximum allowed difference between averaged line slopes of the two line sets. (radians)& 0.15                                                                      \\
 dro                 & Maximum allowed horizontal displacement between line sets. (pixels)                    & 25                                                                        \\
\hline
                     & Detect dim parameters                                                                  &                                                                            \\
\hline
 minFlux             & Pixels with flux less than minFlux are set to zero.                                    & 0.03                                                                       \\
 addFlux             & Pixels with flux larger than zero have addFlux added to them.                          & 0.5                                                                        \\
 erodeKernel         & Erosion kernel.                                                                        & 3x3 matrix, all elements are 1.                                            \\
 dilateKernel        & Dilation kernel.                                                                       & 9x9 matrix, all elements equal to 1.                                       \\
 contoursMode        & Contours mode.                                                                         & RETR\_LIST, return all contours as a flat list.                            \\
 contoursMethod      & Contours approximation method.                                                         & CHAIN\_APROX\_NONE, a full set of points encompassing the contour is returned. \\
 minAreaRectMinLen   & Allowed minimum length of sides of a minimum area rectangle. (pixels)                  & 1                                                                          \\
 lwTresh             & Minimum allowed ratio of sides of a minimum area rectangle.                            & 5                                                                          \\
 nlinesInSet         & Number of detected lines to be declared a set.                                         & 3                                                                          \\
 thetaTresh          & Maximum allowed difference between any two line slopes in a set. (radians)             & 0.15                                                                      \\
 lineSetTresh        & Maximum allowed difference between averaged line slopes of the two line sets. (radians)& 0.15                                                                      \\
 dro                 & Maximum allowed horizontal displacement between line sets. (pixels)                    & 25                                                                        \\
\end{tabular}
\end{table*}

\section{Discussion and conclusion}
\label{sec:DiscussionConclusion}

\subsection{False positives}
\label{subsection:falsepositive}

Our LFDA is highly resistant to false-positive detections, mainly due to its step when it compares lines obtained through minimum-area rectangles with lines from the processed image. The reason why this step is so efficient can be understood if we consider what type of signals tend to emerge as false positives. Those are objects that are not removed or not entirely removed from the image. This is often the case with very bright objects, very dim objects and objects with visible internal structure. Very dim objects and objects with internal structure tend to have poorly determined petrosian radii, which results in a bad masking element. Very bright objects are saturated and their centroids cannot be determined. When this occurs, the masking element is not placed at the appropriate pixel coordinates. Objects that are only partially removed have generally well-defined measured properties, but the objects are bigger than the masking element of \emph{maxxy} in size.

The minimum-area rectangles fitted to large completely unremoved objects are rejected because they fail to comply with the \emph{lwTresh} condition. This enables LFDA to avoid false positives in such cases because the object will not appear in the reconstructed image (which contains only drawings of elongated minimum area rectangles) and, therefore, cannot satisfy a comparison with the possible lines that the Hough algorithm produces in the processed image. In the case of partially removed objects, the minimum-area rectangle fitted to such an object will depend on the shape of the object. Partially removed objects have only their centers masked out, while some outer brightness structure remains. This makes them appear like a hollow disk and, except in the rare cases of large edge-on galaxies, no minimum-area rectangles fitted to such structures can pass the \emph{lwTresh} condition. In some cases the hollowed-out objects can appear like arcs, especially when there are other masked objects in the vicinity that also mask out parts of the hollow disk as well. Such arcs can possibly yield minimum area rectangles, but the Hough algorithm applied to the processed image would not favor such objects (see Appendix \ref{app:HoughEx}). This again leads to the comparison failure between lines from the reconstructed and processed images.

This process rejects the vast majority of false positive detections and makes our LFDA a very robust algorithm. Nonetheless, some false positive may appear, but since the number of detected lines is typically small, such cases can be easily removed by visual inspection in the post-analysis.

\subsection{False negatives}
\label{subsection:falsenegative}

\begin{figure}

\begin{subfigure}{.5\textwidth}
\centering
 \includegraphics[width=\linewidth]{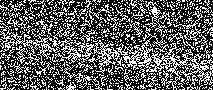}
 \caption{Transparent dim linear features are almost undifferentiated from the surrounding noise. This example is taken from the SDSS {frame-u-005973-3-0127} image frame.}
 \label{fig:zoomin}
\end{subfigure}

\begin{subfigure}{.5\textwidth}
\centering
  \includegraphics[width=\linewidth]{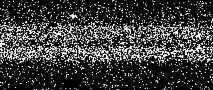}
  \caption{A transparent linear feature exhibiting a defocusing effect visible as a brightness dip along the middle of the line. The interior of the linear feature is full of holes - pixels of very low intensity. This example is taken from the SDSS {frame-g-002728-2-0424} image frame.}
  \label{fig:zoomin2}
\end{subfigure}

  \caption{Two examples of transparent linear features that tend to avoid detection with LFDA. Their low brightness makes them barely visible and only a view from a distance reveals their existence as an increase in white pixel density. The size of kernels used in the LFDA morphological operators is typically fine tuned to the brighter, more pronounced features. This makes them unfit to detect these wide transparent trails. Additionally, not all linear features have a uniform internal structure, as seen in the lower panel example.}
   \label{fig:transZOOM}
\end{figure}

More worrisome are situations when a linear feature is present in the image but is not detected. In such cases we would not know that some lines still exist in the processed sample and our conclusions based on detected lines might become biased. We recognized three classes of linear features that our LFDA has a problem identifying:
\begin{itemize}
\item Cases when the overzealous object removal process removes too much of the line, as shown in Figure \ref{fig:falseneg}.
\item Extremely dim, translucent features (examples are shown in Figure \ref{fig:transZOOM}) are often destroyed during the erosion step. The mechanism of how that happens has been discussed in section \ref{sec:DetectDim}.
\item Lines that have an internal brightness structure in a form of a brightness dip along the middle od the line. This feature is typically a sign of defocusing effect in meteors imaged through big telescopes. It can also be due to meteor fragmentation, when multiple brightness dips might emerge. Such linear features can be very wide, as shown in a 48 pixels wide example in Figure \ref{fig:zoomin2}. The problem with these lines is that they can fail on one of the line consistency checks (\emph{dro}, \emph{linesetTresh} or \emph{thetaTresh}).
\end{itemize}
These shortcomings can be bypassed in certain situations as the sources of long linear features are transient objects and the fields of view are not particularly large. Imaging devices on large scale sky surveys are generally constructed out of tens of CCDs arranged in a square, rectangular or circular arrays in which each CCD records one or multiple images stored into a database as imaging frames. By knowing the geometry of this array the path of a linear feature across the array can be reconstructed from a detection in a single frame. This way we can look for the linear feature even in frames where LFDA fails to detect the line.

\subsection{Implementation of LFDA}
\label{subsection:implementation}

During the development phase we tested our LFDA on a large variety of images, mostly from the SDSS database. In general, almost all trails could have been detected with a careful selection of the detection parameters. We see that our LFDA performs fast enough, 0.1 to 0.3 seconds per image frame, in small scale tests \citep{LSSTEurope}. We will soon finish a large scale test on the entire SDSS database and reveal the results in the upcoming publication. It will cover the questions of algorithm performance, possible existing execution bottlenecks, selection of optimal detection parameter values and corresponding detection rates for the entire SDSS database. For the purpose of measuring detection rates we identified close to complete set of frames with lines in the SDSS $r$ and $i$ filters (our high degree of confidence in the completeness arises from inspecting by hand more than 500 000 images that were suspected for containing linear features) and tests show 80\% and 74\% detection rates, respectively \citep{LSSTEurope}. So far we see that our LFDA is fast, robust, with low false positives/negatives, easily parallelizable, and easily adaptable to different computational environments.

The applicability of LFDA in diverse observational settings is of a great concern. Apparat from the required modifications to the object removal step and the optimization of detection parameter values, no other part of LFDA would have to be changed for implementation in different environments. The method does not require a large scale testing to derive optimal algorithm parameter values. This can be done on a small initial test set, which enables its easy implementation into the image processing pipelines of existing and upcoming sky surveys. The source code of our LFDA is available at
\url{https://github.com/DinoBektesevic/LFDA}
under the General Public License GPLv3\footnote{\url{https://www.gnu.org/licenses/gpl-3.0.en.html}}.

Sky surveys are particularly suitable for LFDA as they already have an object detection pipeline, which is a necessary component of LFDA within its initial object removal step. High cadence sky surveys represent an especially attractive target for LFDA implementation because they use image differencing in their data pipelines. This approach can be almost perfect in its ability to detect all the objects in an image, which removes a lot of obstacles that cause false positive or false negative detections. However, implementations of LFDA are not limited to sky surveys alone. As mentioned in section \ref{sec:objectremoval}, there are other ways by which the object removal step can be achieved. The most general, although one of the slowest, would be applying SExtractor to detect and remove objects such as stars and galaxies. Hence, we see our LFDA as a useful tool for any research aiming at identification of long linear features in large astronomical imaging datasets.

In the upcoming publication we will address the issue of meteor defocusing that affects the shape of the lines (Figure \ref{fig:zoomin2}), but enables differentiation between satellites and meteors, as well as extraction of meteor physical properties from the track brightness profile. This will be followed by a publication that describes results of a large scale test on the entire SDSS database, which will also enable us to publish a large set of SDSS frames with lines. This set can be used in the future for testing new linear feature detection algorithms, including machine learning methods that need training sets. The goal is to reach a level of automatization where algorithms not only detect lines, but also classify them based on their origin and extract some physical properties of the moving object.

\section*{Acknowledgements}

We thank Darko Jevermovi\'{c} for his help with the SDSS database setup and access to Fermi cluster at the Observatory Belgrade. We are thankful to Aleksandar Cikota for helpful discussions and a preliminary investigation into this topic. We acknowledge the help by Benjamin A. Weaver from the Physics Department, New York University, for his help with the SDSS data. Authors are also thankful to Russ Laher for usefull comments on our manuscript. We also thank the Physics Department, University of Split, and the Technology Innovation Centre Me\dj{}imurje for additional computational resources.

The authors acknowledge being apart of the network supported by the COST Action TD1403 Big Data Era in Sky and Earth Observation, which helped with discussions on the applicability of our research.

We would also like to thank Sloan Digital Sky Survey for the free and open data access and accompanying detailed documentations. Funding for SDSS-III has been provided by the Alfred P. Sloan Foundation, the Participating Institutions, the National Science Foundation, and the U.S. Department of Energy Office of Science. The SDSS-III web site is http://www.sdss3.org/.

SDSS-III is managed by the Astrophysical Research Consortium for the Participating Institutions of the SDSS-III Collaboration including the University of Arizona, the Brazilian Participation Group, Brookhaven National Laboratory, Carnegie Mellon University, University of Florida, the French Participation Group, the German Participation Group, Harvard University, the Instituto de Astrofisica de Canarias, the Michigan State/Notre Dame/JINA Participation Group, Johns Hopkins University, Lawrence Berkeley National Laboratory, Max Planck Institute for Astrophysics, Max Planck Institute for Extraterrestrial Physics, New Mexico State University, New York University, Ohio State University, Pennsylvania State University, University of Portsmouth, Princeton University, the Spanish Participation Group, University of Tokyo, University of Utah, Vanderbilt University, University of Virginia, University of Washington, and Yale University.








\appendix

\section{Hough transform examples}
\label{app:HoughEx}

\begin{figure}
  \begin{centering}
  \includegraphics[width=0.49\textwidth]{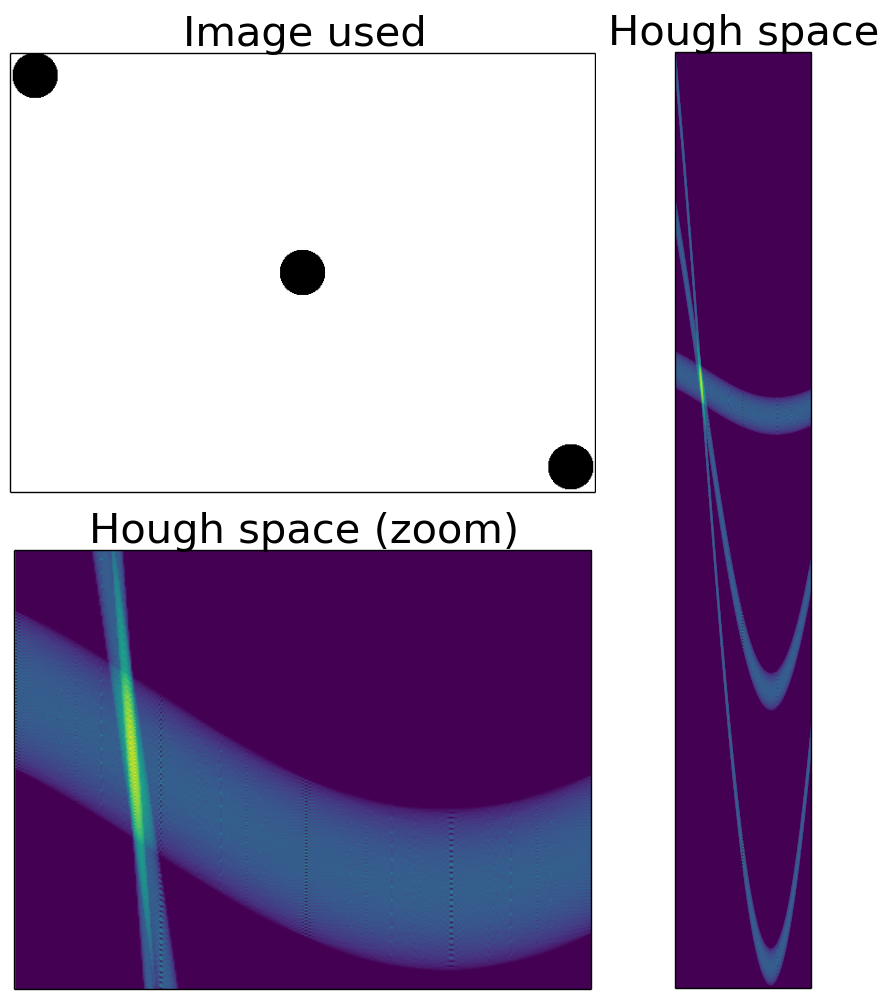}
  \caption{Hough space of an image with three dots and no line.}
  \label{fig:houghdots}
  \end{centering}
\end{figure}

\begin{figure}
  \begin{centering}
  \includegraphics[width=0.49\textwidth]{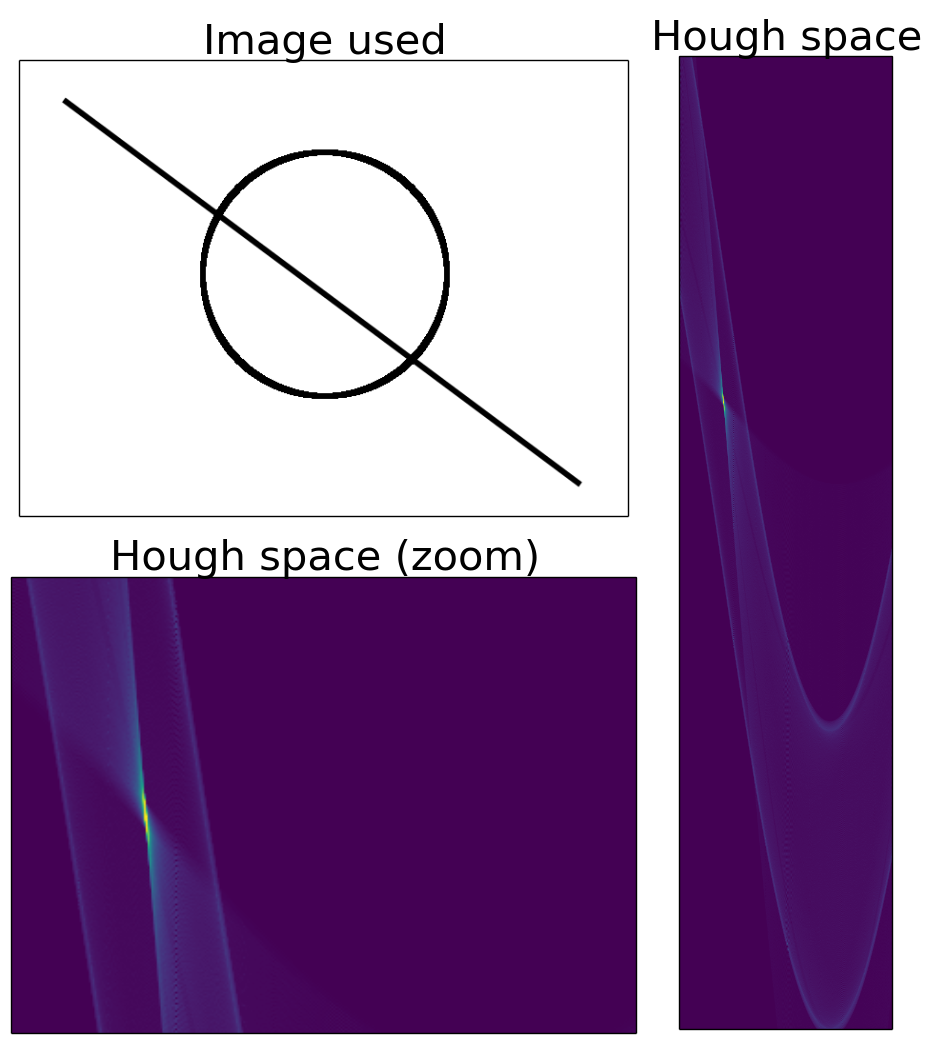}
  \caption{Hough space of an image with a circle and a line.}
  \label{fig:houghline}
  \end{centering}
\end{figure}

\begin{figure}
  \begin{centering}
  \includegraphics[width=0.49\textwidth]{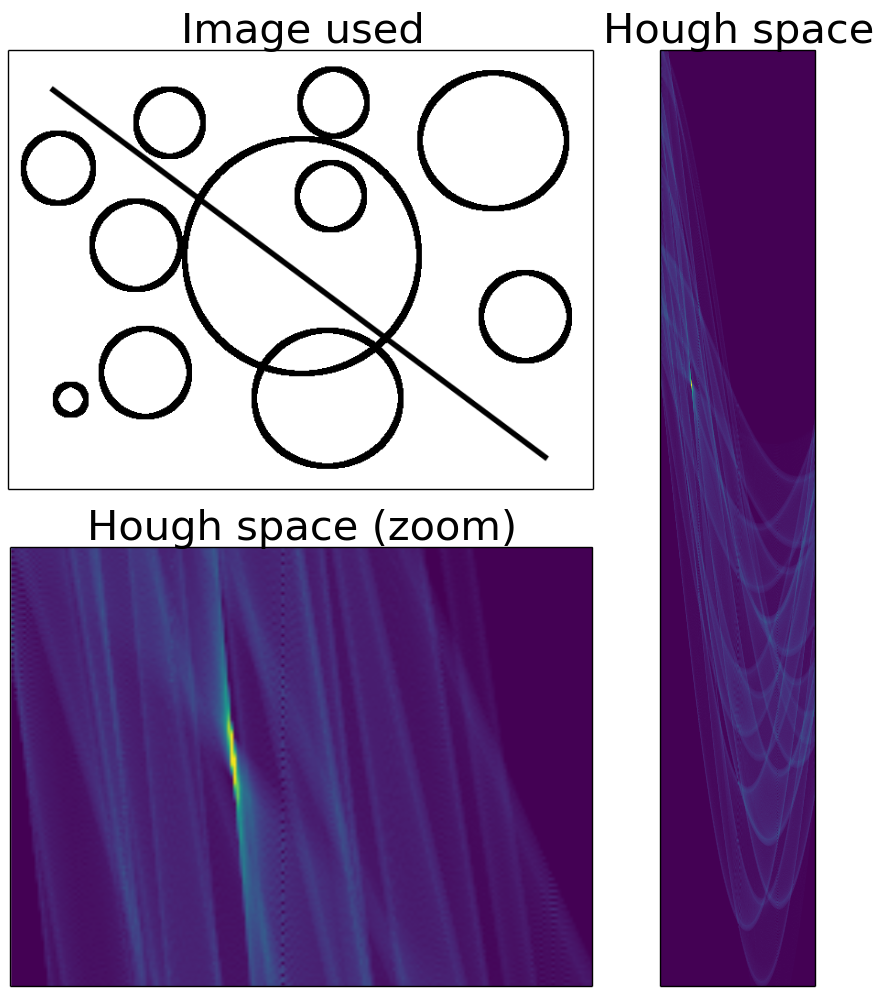}
  \caption{Hough space of an image with a line and a lot of circles.}
  \label{fig:houghcircles}
  \end{centering}
\end{figure}

Figure \ref{fig:houghdots} shows why a careful determination of the position of the accumulator maximum in the Hough $(r, \theta)$  space is so important. A large number of highly active accumulators (i.e. values close to the maximum) are visible in the neighborhood of the detection point (the maximum value in the Hough space) in the zoomed panel in Figure \ref{fig:houghdots}. Each of those active accumulators represents a line that passes through the three dots in the used image with varying precision. A proper centroid location method applied to the maximum neighborhood would probably yield the best $(r, \theta)$ values, but it would also be computationally more demanding. Considering how symmetrically spaced the active accumulators are around the actual value, a good approximative approach would be averaging a set of top $n$ active accumulators (i.e. the first $n$ maxima in the Hough space). Notice also how dots produce pronounced and tight sinusoids in the Hough space compared to the sinusoids produced by circles shown in Figures \ref{fig:houghline} and \ref{fig:houghcircles}. The reason why dots produce such sinusoids is that they are disks. Within disks and disk-like objects there are a lot of neighboring active pixels through which a new line can be drawn. All those lines produce a similar maximum in the Hough space and therefore leave a very distinguishable compact track. It is important to notice that even though a line does not actually exist in Figure \ref{fig:houghdots}, a strong maximum in the Hough space implies a line anyhow.

Figures \ref{fig:houghline} and \ref{fig:houghcircles} show the opposite case of a disk. Unlike the filled dots, circles and generally objects that are not filled produce wider and less distinguishable tracks in the Hough space. This means that all disk-like objects will have to be "hollowed" out, or completely removed, otherwise the Hough line detection algorithm would likely produce a detection, as was the case in Figure \ref{fig:houghdots}. This is why masking objects with squares in the object removal step is successful, provided enough of the objects' interior is masked out. Figure \ref{fig:houghcircles} also shows how robust and unaffected by noise the Hough transform is. The line accumulators in the Hough space remained as visible as in the case of a single circle (Figure \ref{fig:houghline}) even though the used image was far more crowded.

\section{Detailed algorithm flowchart}
\label{sec:detailedalgflow}

\begin{figure}
  \centering
  \includegraphics[width=0.46\textwidth]{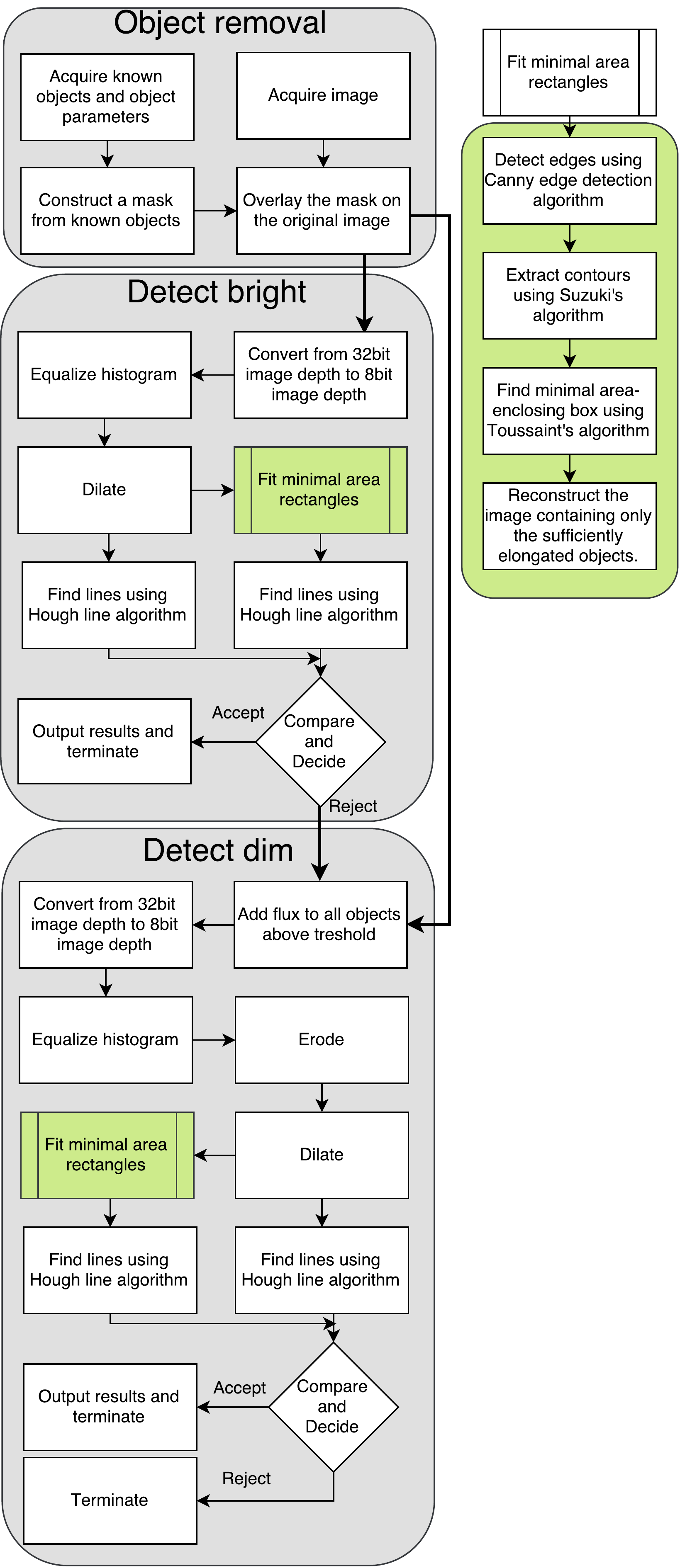}
  \caption{Flowchart diagram describing a detailed algorithm flow.}
  \label{fig:DetailedFlowChart}
\end{figure}

In pseudo-codes \ref{alg:detecttrails} and \ref{alg:createjobs} we show a general logic flow of our LFDA, while here in Figure \ref{fig:DetailedFlowChart} we show in more detail the algorithm flow of the key parts - object removal (in case images do not already have objects removed), detecting bright linear features and detecting dim linear features.

%
%

%
%

%


\bsp	
\label{lastpage}
\end{document}